\documentclass[11pt]{article}
\usepackage{graphicx}
\usepackage{amsmath}
\usepackage{amssymb}
\usepackage{mathrsfs} 
\usepackage{amsthm}
\usepackage{tikz}
\usepackage{pgfplots}
\usetikzlibrary{intersections, pgfplots.fillbetween, shapes, arrows,decorations.pathmorphing, patterns}
\tikzstyle{snakeline} = [decorate, decoration={pre length=0.1cm,
                         post length=0.1cm, snake, amplitude=.5mm,
                         segment length=2mm},thick, red, ->]
\usepackage{subcaption}

\usepackage{hyperref}
\hypersetup{
    colorlinks=true,
    linkcolor=black,
   citecolor=blue,
   urlcolor=black,
}

\usepackage{authblk}
\usepackage[text={6in,9in},centering]{geometry}
\numberwithin{equation}{section}

\newcommand{\pa}{\partial}

\newcommand{\Tr}{{\rm Tr}}
\newcommand{\Lie}{{\pounds}}

\renewcommand{\a}{\alpha}
\renewcommand{\b}{\beta}
\renewcommand{\c}{\gamma}

\renewcommand{\d}{\delta}

\newcommand{\s}{\sigma}
\newcommand{\Si}{\Sigma}
\renewcommand{\t}{\theta}

\newcommand{\la}{\lambda}

\newcommand{\ud}{\mathrm{d}}          
\newcommand{\ue}{\mathrm{e}}    
\newcommand{\be}{\begin{equation}}
\newcommand{\ee}{\end{equation}}
\newcommand{\bea}{\begin{eqnarray}}
\newcommand{\eea}{\end{eqnarray}}
\newcommand{\ba}{\begin{align}}
\newcommand{\ea}{\end{align}}

\newcommand{\nn}{\nonumber\\}

\newcommand{\cL}{\mathcal{L}}
\newcommand{\cN}{\mathcal{N}}
\newcommand{\cM}{\mathcal{M}}
\newcommand{\cH}{\mathcal{H}}
\newcommand{\cO}{\mathcal{O}}
\newcommand{\cF}{\mathcal{F}}
\newcommand{\cA}{\mathcal{A}}

\newcommand{\cC}{\mathcal{C}}
\newcommand{\cD}{\mathcal{D}}

\newcommand{\cK}{\mathcal{K}}
\newcommand{\cZ}{\mathcal{Z}}
\newcommand{\cW}{\mathcal{W}}

\newcommand{\cS}{\mathcal{S}}

\title{ 
Semiclassical $p$-branes in hyperbolic space 
\vspace{8mm}}

\author{\normalsize  Rodrigo de Le\'on Ard\'on\thanks{ \href{rdeleon@ecfm.usac.edu.gt}{rdeleon@ecfm.usac.edu.gt}
}\vspace{4mm}}

\affil{ \normalsize \it Departamento de Ciencias F\'isicas, \\
\it Facultad de Ciencias Exactas, \\
\it Universidad Andr\'es Bello,\\
\it Sazi\'e 2212, Piso 7, Santiago, Chile.}

\affil{ \normalsize \it Instituto de Investigaci\'on en Ciencias F\'isicas y Matem\'aticas,\\ 
\it Escuela de Ciencias F\'isicas y Matem\'aticas,\\
\it Universidad de San Carlos de Guatemala,\\
\it Ciudad Universitaria, Zona 12 Guatemala.}

\date{}
\begin{document}
\maketitle

\begin{abstract}
The one-loop effects to the Dirac action of $p$-branes  in a hyperbolic background from the path integral and the solution of the Wheeler-DeWitt equation are analysed. The objective of comparing the equivalent quantization procedures is to study in detail the validity of the semiclassical approximation and divergences associated to one-loop corrections. This is in line with a bottom-up approach to holographic Wilson loops.
We employ the heat kernel regularization method for both quantization procedures and we study in great detail one-loop corrections to geodesics in a 2-dimensional hyperbolic space and semi-spheres in a 3-dimensional hyperbolic space. We show that the divergences, given by the high energy expansion of the heat kernel, can be classified by their compatibility with the semiclassical approximation and geometric nature.
\end{abstract}

\newpage
\tableofcontents


\section{Introduction}
In this note, we study one-loop effects to the Dirac action of $p$-branes in a hyperbolic background from the path integral approach and the solution of the Wheeler-DeWitt equation. This is in line with a bottom-up approach  to the AdS/CFT conjecture proposed in \cite{Maldacena:1997re}. In particular, this study is motivated by the computations of leading order corrections to holographic Wilson loops from the gravity perspective, see \cite{Kinar:1999xu, Forste:1999qn, Drukker:2000ep, Kruczenski:2008zk, Beccaria:2010zn,Faraggi:2011bb,Faraggi:2011ge,Kristjansen:2012nz,Kim:2012tu,Forini:2014kza,Buchbinder:2014nia,Forini:2015mca,Bergamin:2015vxa,Faraggi:2016ekd, Forini:2017whz,Cagnazzo:2017sny,Chen-Lin:2017pay,Medina-Rincon:2018wjs,Aguilera-Damia:2018bam,Medina-Rincon:2019bcc,David:2019lhr,Hernandez:2019huf}. 

The objective of comparing the equivalent quantization procedures is to study in detail the validity of the semiclassical approximation and divergences associated with one-loop corrections. We compute the semiclassical quantization of the  extended objects in a hyperbolic space and  consider the Poincar\'e half plane model of such  background. This choice of coordinate system allows us to discuss near the boundary phenomena in full detail.

 In the canonical approach, reparametrization invariance of the Dirac action implies that the Hamiltonian $H$ is constrainted to vanish, see for example \cite{Henneaux:1992ig}. The quantum version of this constraint, in Euclidean signature, gives the Euclidean Wheeler-DeWitt equation
\be
H\Psi=0,
\ee 
where $\Psi$ corresponds to the Euclidean wavefunctional. For the WKB approximation  we write $\Psi=\exp\left(-\cS/\hbar\right)$ and assume that  $\cS$ admits an expansion of the form $\cS=\cS_{(0)}+\cS_{(1)}\hbar+\ldots$. In the limit $\hbar\to 0$ the resulting equation corresponds to the Euclidean Hamilton-Jacobi equation and $\cS$ is identified with Euclidean Hamilton's principal function. As stated in \cite{Halliwell:1990uy}, a particular solution of the Wheeler-DeWitt equation is obtained by imposing boundary conditions on $\Psi$. This in turn implies that we obtain  a particular solution of $\cS$  in the WKB approximation. Boundary conditions on the wave functional corresponds to initial conditions on Hamilton's principal function. In close analogy to the approaches of computing the wavefunction of the universe in quantum cosmology, see for example \cite{Hawking:1980gf,Vilenkin:1982de,Hartle:1983ai,Vilenkin:1986cy,Vilenkin:1987kf,Halliwell:1988wc,Halliwell:1989dy,Halliwell:1990uy},  we also consider the path integral representation of the wavefunctional. It is written as
\be
\Psi = \int\cD\phi^i\,\ue^{{-S_{\mathrm{Dirac}}[\phi^i]/\hbar}}, 
\ee
where $S$ is the Euclidean Dirac action, $\{\phi^i(x)\}$ is the set of fields that form the coordinate system of the background and  $\{x\}$ parametrizes the extended object\footnote{In some sense, the Dirac action can be thought as a kind of minisuperspace model with only matter variables, see \cite{Halliwell:1988wc,Halliwell:1990uy}. The minisuperspace metric corresponds to the background metric. This realization is more transparent if instead of the Dirac action we consider the (classically equivalent) Polyakov action. }. The boundary conditions for the path integral must be chosen  adequately for the integral to satisfy the Wheeler-DeWitt equation. The semiclassical limit of the path integral is given by the saddle point and it is of the form
\be
 \Psi \approx \ue^{-S_{\mathrm{Dirac}}^{*}/\hbar}
 \int\cD\zeta^i \,\,\ue^{-S_{\mathrm{Dirac}}^{(2)}/\hbar}\approx\frac{\ue^{-S_{\mathrm{Dirac}/\hbar}^{*}}}{\sqrt{\det\cO'}},
\ee
where $S_{\mathrm{Dirac}}^{*}$ is the Dirac action on-shell, $S_{\mathrm{Dirac}}^{(2)}$ is the quadratic action for the fluctuating fields $\zeta^i$ and $\cO$ is the operator of the fluctuations. The prime in $\det\cO'$ indicates that we have removed zero modes. Note that we have chosen a gauge that gives a purely local Fadeev-Popov determinant. Dirichlet  boundary conditions for the path integral of the fluctuations is the natural choice for objects with fixed endpoints. Other boundary conditions are of course allowed and they may be required for supersymmetric models and backgrounds with more complicated topology. 

We see that the one-loop correction computed from both methods implies
\be
\ue^{-\cS_{(1)}}\Leftrightarrow\frac{1}{\sqrt{\det\cO'}}. 
\ee
The divergences from both points of view can be compared. Let $\{\la_n\}$ be the spectrum of $\cO$, then
\be
\cS_{(1)}\to \infty\Leftrightarrow  \prod_{n}^{\infty }\phantom{}^{'} \la_n\to\infty,\quad \cS_{(1)}\to -\infty\Leftrightarrow  \prod_{n}^{\infty }\phantom{}^{'}\la_n\to 0.
\ee
In the canonical approach the $\cS_{(1)}\to -\infty$ divergence is related to the validity of the semiclassical approximation. This can be seen by recalling that in the one-dimensional case  $\exp(-\cS_{(1)})$ diverges to infinity at the turning points of the potential. Therefore, the WKB approximation is only valid far away from these points. 

In order to handle  divergences in a systematical way and all encompass view, we follow \cite{Voros:1986vw,voros1992,jorgenson1993basic}. Where they discuss the regularization of the functional determinant via the zeta function associated to an operator. This in turn implies a relation with the trace of the heat kernel (the basic theta-type function) of this operator. As stated in \cite{voros1992}, the trace of the heat kernel contains more spectral information since it bridges (via integral transforms) functional determinants with zeta functions. The heat kernel by itself is a useful object for calculating quantum fluctuations since among many  virtues, we can highlight that it is a fully covariant object, see for example \cite{Vassilevich:2003xt,Fursaev:2011zz}. 

In the context of extended objects in AdS/CFT, regularization methods have been studied in \cite{Forini:2017ene,Aguilera-Damia:2018rjb, Aguilera-Damia:2018twq} and the heat kernel method has been employed in \cite{Drukker:2000ep,Faraggi:2011ge,Buchbinder:2014nia,Bergamin:2015vxa,Forini:2017whz,Aguilera-Damia:2018bam}. Unfortunately this method does not necessarily gives the correct answer to the one-loop correction of the partition function after it is compared with the result from the field theory side. The discrepancy may arise from subtleties inherent of each problem\footnote{These subtleties may arise from zero modes. This implies that one must  integrate over these modes as dictated by the collective coordinate method, see for example \cite{Gervais:1974dc,Bernard:1979qt,Dorey:2002ik,Tong:2005un}. This will be discussed in section 3.} and signals the necessity of a more detailed study of the high energy expansion of the heat kernel. In particular, the relation of the heat kernel  with the phase shift method needs clarification as the later method  has been succesfully employed in \cite{Chen-Lin:2017pay,Medina-Rincon:2018wjs,Medina-Rincon:2019bcc}.

In view of these issues, this note is organized as follows. In section 2, we review the geometry of $p$-branes and their fluctuations. We discuss the one-loop correction from both methods of quantization in section 3. For the path integral method, we include a brief discussion of zero modes. We also compute the one-loop correction to the Wheeler-DeWitt equation in the hyperbolic background. This equation is regulated using the heat kernel  and we set the conditions that makes the WKB approximation valid. In section 4, we discuss in full detail a simple but non-trivial example: the spectrum of fluctuations of a two-dimensional worldline. The relation of this spectrum  with the  semiclassical wavefunction of the worldline via the Wheeler-DeWitt equation and regularization methods of the one-loop corrections is discussed. We show how the heat kernel method encompasses other methods such as phase-shifts and with the aid of an auxiliary statistical mechanical system we give a general framework to compute the quantum corrections.

In section 5, we apply the methods to a toy scenario of holographic Wilson loops in Euclidean AdS$_3$. We briefly review the work done in these lines. We focus on the circular Wilson loop and discuss the relation between divergences encountered by the semiclassical Wheeler-DeWitt solution and the heat kernel. We provide a relation between the regularization schemes involved. The density of states and zero modes are analysed. Finally in section 6, we conclude and provide a discussion of a possible classification of the divergences.

\section{Geometrical setup}
\subsection{A review on fluctuations of $p$-branes}
A classical scalar field configuration is defined as the differentiable map $\phi:(\Si,h)\to(\cM,g)$ from a $D$-dimensional (pseudo) Riemannian manifold $(\Si,h)$ to a $d$-dimensional (pseudo) Riemannian manifold $(\cM,g)$, which is referred to as the target space. Let us consider the chart $(U,\psi_1)$ in $\Si$ and the chart $(V,\psi_2)$ in $\cM$ together with the points $p\in U$ and $f(p)\in V$. The map $\psi_2\circ \phi\circ \psi_1^{-1}$ with $\psi_1(p)=\{x^{\mu}\}$ and $\psi_2(\phi(p))=\{y^i\}$ corresponds to the components $y^i=\phi^i(x^{\mu})$ of the scalar field. 

If $\dim \Si < \dim\cM$, we can further interpret $\phi(\Si)$ as a submanifold of $\cM$. Then $\phi:\Sigma\to\cM$ defines an isometric embedding. $\cM$ becomes the ambient space and the coordinates $\phi^i(x^{\mu})$ describe a submanifold of codimension $\dim\cM-\dim\Si$. For convenience, we set $\dim\Si=1+p$ and refer to the $p$-coordinates as ``spatial''. For $p=0$, $\phi^i(x^0)$ corresponds to a curve in $\cM$. For $p=1$, $\phi^i(x^0,x^1)$ corresponds to a 2-dimensional surface and therefore $\phi^i(x^0,x^1,\ldots,x^p)$ corresponds to a $p$-dimensional surface referred to as $p$-membrane or just simply $p$-brane. The manifold $\Si$ is  interpreted as a mathematical space of parameters. The dynamics of the $p$-brane is given by the Dirac action 
\be
S_{\mathrm{Dirac}} =-M_{(p)}\int\ud^{1+p} x \sqrt{|\det(\phi^*g)|}=-M_{(p)}\int\ud^{1+p}x \sqrt{|\det\left(g_{ij}(\phi)\pa_{\mu}\phi^i\pa_{\nu}\phi^j\right)|},
\label{Dirac}
\ee
where $M_{(p)}$ has dimensions of (mass)$^{1+p}$. The equations of motion derived from this action are 
\be
 K^i(\phi)\equiv \Box\phi^i+\Gamma_{lm}^i(\phi)D_{\mu}\phi^lD^{\mu}\phi^m=0,
 \label{eom}
\ee
where $\Box=D_{\mu}D^{\mu}$ and $D_{\mu}$ are the Laplacian and the covariant derivative with respect to the induced metric respectively, $\Gamma_{ijk}(\phi)$ are the Christoffel symbols constructed from the target space metric and we have assumed the orthogonality condition $g_{ij}(\phi)\pa_{\mu}\phi^i\d\phi^j=0$.

At the quantum level, the quantization via the path integral of the Dirac action is difficult basically due to the non-linear nature of the action and therefore, we consider Polyakov action. This action is defined with the aid of the auxiliary field $h_{\mu\nu}$ as
\be
S_{\mathrm{Polyakov}}[\phi^i,h]=-\frac{M_{(p)}}{2}\int_{\Si}\ud^{1+p} x\sqrt{|\det h|}\left[g_{ij}(\phi)h^{\mu\nu}\pa_{\nu}\phi^i\pa^{\mu}\phi^j-(p-1)\right]. 
\label{Asc2}
\ee
Plugging the equations of motion for $h^{\mu\nu}$, we obtain the Dirac action. At the classical level the actions given by Eq. \eqref{Asc2} and Eq. \eqref{Dirac} are equivalent. The field $h$ in  Eq. \eqref{Asc2} and the metric $\phi^*g$ in Eq. \eqref{Dirac} are in fact the same  for $ p\neq 1$ (or $D\neq 2$) or otherwise, the former differ from the later by a conformal factor. This is due to the invariance of Eq. \eqref{Asc2} under Weyl transformations of the metric $h$ for $p=1$.

Let us investigate the quadratic part of the Taylor expansion of the action given in Eq. \eqref{Dirac}. In order to do so we follow \cite{AlvarezGaume:1981hn} and  introduce  the Riemann normal coordinate system. Let us consider the parametric curve $\c : [a,b]\to\cM$ of unit length and the function
\be
 s(t)=\int\limits_a^t\ud u\, \|\c'(u)\|,
\ee
with $t\in[a,b]$. Consider $s(t)$ to be the parameter of the curve and let the two neighbouring points $O$ and $P$ be joined by the curve $\c$ in the ambient space. In a local coordinate system, the curve is given by $\{\xi^i(x;s)\}$. Let $O=\{\xi^i(x;0)\}$,  $P=\{\xi^i(x;1)\}$ and consider 
\be
\xi^i(x;s)=\xi^i(x;0)+\left.\pa_s \xi^i(x;s)\right|_{s=0} s+\frac{1}{2}\left.\pa^2_s \xi^i(x;s)\right|_{s=0} s^2+O(s^3).
\ee 
Thus, the tangent vector at $O$ is $\left.\pa_s \xi^i(x;s)\right|_{s=0} $. Let us further assume that the curve $\c$ is actually a geodesic such that it satisfies
\be
\pa^2_s \xi^i(x;s)+\Gamma^i_{lm}\pa_s \xi^l(x;s)\pa_s \xi^m(x;s)=0,
\ee
This implies
\be
\phi^i(x)=\bar{\phi}^i(x)+\zeta^i(x)-\frac{1}{2} \Gamma^i_{jk}(\bar\phi)\zeta^j(x)\zeta^k(x)+O(\zeta^3),
\label{paphi1}
\ee 
where $\phi^i(x)=\xi^i(x;1)$, $\bar{\phi}^i(x)=\xi^i(x;0)$ and  $\zeta^i(x)=\zeta^i(\bar\phi(x))=\left.\frac{\pa \xi^i(x;s)}{\pa s}\right|_{s=0}$. The fields $\bar{\phi}^i(x)$ are called the background fields and $\zeta^i(x)$, which correspond to vector field components in the ambient space, are the fluctuation fields. 
If furthermore the geodesics satisfy the condition $\Gamma^i_{lm}(\bar\phi)=0$, the coordinate system at $P=\{\phi^i(x)\}$ is referred as to Riemann normal coordinate system. In this coordinate system we have $\left.\pa _lg_{ij}\right|_{\phi=\bar{\phi}}=0$  and
\be
g_{ij}(\phi)= g_{ij}(\bar{\phi})-\frac{1}{3}R_{iljm}\zeta^l\zeta^m+O(\zeta^3),
\ee
where the Riemann tensor is evaluated at $P$. The pullback metric in this coordinate system results in\footnote{In order to derive this result, first compute the expansion using \eqref{paphi1} and from the final result apply the Riemann normal coordinates properties.}
\begin{multline}
\phi^*(g)_{\mu\nu}= g_{ij}(\bar\phi)\pa_{\mu}\bar\phi^i\pa_{\nu}\bar\phi^j+2\pa_{(\mu}\left[g_{ij}(\bar\phi)\pa_{\nu)} \bar\phi^i \zeta^j\right]-2\zeta^i g_{ij}(\bar\phi)\kappa^j_{\mu\nu}(\bar{\phi})\\+g_{ij}(\bar\phi)\pa_{\mu}\zeta^i\pa_{\nu}\zeta^j-R_{limj}\pa_{\mu}\bar\phi^l\pa_{\nu}\bar\phi^m\zeta^i\zeta^j+O(\zeta^3),
\label{flucinducedmetric}
\end{multline}
where $\kappa^i_{\mu\nu}(\bar{\phi})$ is for the moment a book keeping device defined, in a general coordinate system in the target space, as
\be
\kappa^i_{\mu\nu}(\bar{\phi})\equiv \pa_{\mu}\pa_{\nu}\bar\phi^i+\Gamma^i_{lm}(\bar\phi)\pa_{\mu}\bar\phi^l\pa_{\nu}\bar\phi^m.
\ee
In Riemann normal coordinates, we have $\pa_{\mu}\zeta^i=\pa_{\mu}\bar\phi^l\pa_l\zeta^i=\pa_{\mu}\bar\phi^l\bar\nabla_l\zeta^i$ where  $\bar\nabla$ is the covariant derivative at $O$. Therefore, $\pa_{\mu}$ acting on the fluctuations corresponds to the pullback of the covariant derivative of $\bar\nabla$.

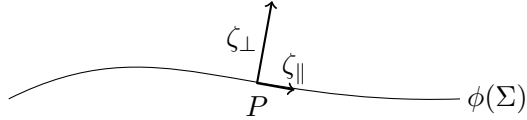
\begin{figure}[ht!]
\centering
\begin{tikzpicture}
\draw (0,0) .. controls (2,1) and (3,-0.1) .. (6,0);
\node at (6.5,0) {$\phi(\Si)$};
\draw[->,thick] (3.3,0.2) -- (3.5,1.3);
\draw[->,thick] (3.3,0.21) -- (3.8,0.12);
\node at (3.3,-0.1) {$P$};
\node at (3.1,0.8) {$\zeta_{\perp}$};
\node at (3.8,.4) {$\zeta_{\parallel}$};
\end{tikzpicture}
\caption{Decomposition of the fluctuations on $\cM$.}
\label{DFpic}
\end{figure}

With the  purpose of simplification and understanding the geometry behind Eq. \eqref{flucinducedmetric}, let us consider $v$ to be a tangent vector of $\Si$ defined at the point $\phi^{-1}(O)$ with $O\in\phi(\Si)\subset\cM$. We can pushforward this vector into the target space and the corresponding vector is $v^i\pa_i= v^{\mu}\pa_{\mu}\bar{\phi}^i\pa_i$. The vector $v^i\pa_i$ at $O$ is tangent to $\phi(\Si)$. On the other hand, a vector $V$ defined in $\cM$ at $O$ admits the decomposition $V = V_{\parallel}+V_{\perp}$. The vector $V_{\perp}$ is perpendicular or normal to $\phi(\Si)$,  i.e. $g_{ij}(\bar\phi) \pa_{\mu}\bar\phi^i V^j_{\perp}=0$. After identifying $v^i\pa_i$ with $V_{\parallel}$, the vector $V$ is referred to as the extension of $v$. Therefore, the vector field $\zeta$ is decomposed into tangent and normal components with respect to $\phi(\Si)$, see Figure \ref{DFpic}. 
In the  expansion given in Eq. \eqref{flucinducedmetric}, it is rather convenient to assume that the fluctuations are purely normal. It is not only due  to the presence of the total derivative but also because the tangential and normal components of the fluctuations mix with each other at second order in the expansion.  The simplest justification for this assumption is that corresponds to a characteristic of the coordinate system. In the literature this gauge choice is referred to as the ``normal'' gauge, see \cite{FrancoisMS}. At the quantum level, the ``normal'' gauge enjoys the feature that the Fadeev-Popov determinant is purely local\footnote{As stated in \cite{FrancoisMS}, the normal gauge condition is defined as $G=g_{ij}(\bar{\phi})\pa_{\mu}\bar{\phi}^i\zeta^j$. Under a diffeomorphism $x^{\mu}\to x^{\mu}-\eta^{\mu}(x)$, a scalar field transforms infinitesimally as $\phi^i(x)\to \phi^i(x)+\eta^{\mu}\pa_{\mu}\phi^i(x)$, and
\bea
\d G &=& g_{ij}(\bar{\phi})\pa_{\mu}\bar{\phi}^i\d\zeta^j,\nn
&=&g_{ij}(\bar{\phi})\pa_{\mu}\bar{\phi}^i\d\phi^j,\quad (\zeta^i=\phi^i-\bar\phi^i),\nn
&=&\eta^{\nu}g_{ij}(\bar{\phi})\pa_{\mu}\bar{\phi}^i\pa_{\nu}\phi^j,\nn
&=&\eta^{\nu}\left[g_{ij}(\bar{\phi})\pa_{\mu}\bar{\phi}^i\pa_{\nu}\bar\phi^j+g_{ij}(\bar{\phi})\pa_{\mu}\bar{\phi}^i\pa_{\nu}\zeta^j\right].\nonumber
\eea
Hence, the Fadeev-Popov determinant is
\be
 \Delta_{FP}=\det\left[g_{ij}(\bar{\phi})\pa_{\mu}\bar{\phi}^i\pa_{\nu}\bar\phi^j+g_{ij}(\bar{\phi})\pa_{\mu}\bar{\phi}^i\pa_{\nu}\zeta^j\right].
\ee
Using the equations of motion of the background field and the properties of Riemann normal coordinates of the ambient space, we find $g_{ij}(\bar{\phi})\pa_{\mu}\bar{\phi}^i\pa_{\nu}\zeta^j=0$ after differentiating by parts.
This corresponds to the Monge representation of the submanifold or the static gauge in string theory literature. Considering the path integral over the fluctuations, the Fadeev-Popov determinant for the ``normal'' gauge is independent of the fluctuations and consequently, the ghosts do not couple with them and the Fadeev-Popov determinant can be absorbed by the path integral normalization constant. }. 
We investigate further consequences of this gauge choice. Let us define the unit normal one-form $N_i$ with $g^{ij}(\bar\phi)N_iN_j\equiv\varepsilon=\pm 1$ and the projectors
\be
\Pi_{ij}^{\parallel} +\Pi_{ij}^{\perp}=g_{ij}(\bar\phi),\quad \Pi_{ij}^{\perp}=\varepsilon N_iN_j.
\ee
The projector $\Pi_{ij}^{\parallel}$ is referred as to the \emph{first fundamental form} while the \emph{second fundamental form} or \emph{extrinsic curvature} is defined as $\cK_{ij}=1/2\Lie_N \Pi_{ij}^{\parallel}$. After computing the Lie derivative we obtain
\be
\cK_{ij}= \cD_{(i} N_{j)},
\ee
where $\cD_i N_j \equiv  (\bar\nabla_i -\varepsilon A_i)N_j$ with $A_i=\Lie_N N_i$ and $A_iN^i=0$. In geometrical terms, $A_i$ corresponds to the acceleration of the integral curves of $N_i$ (recall that the norm of $N_i$ is constant).

An important feature of the extrinsic curvature is that is orthogonal to the normal direction, i.e. $N^i\cK_{ij}=0$. Since the normal vector depend on the $\bar{\phi}^i$ coordinates, the pullback of the extrinsic curvature gives
\be
\cK_{\mu\nu}=\pa_{\mu}\bar\phi^i\pa_{\nu}\bar\phi^j\cK_{ij}=-N_l\kappa^l_{\mu\nu}(\bar\phi).
\ee
After identifying $N_i$ as the dual of $\zeta^i_{\perp}$ with respect $g_{ij}(\bar\phi)$, the expansion of the pullback  metric up to second order becomes
 \be
\phi^*(g)_{\mu\nu}= g_{ij}(\bar\phi)\pa_{\mu}\bar\phi^i\pa_{\nu}\bar\phi^j+2\cK_{\mu\nu} +g_{ij}(\bar\phi)\pa_{\mu}\zeta^i_{\perp}\pa_{\nu}\zeta_{\perp}^j-R_{limj}\pa_{\mu}\bar\phi^l\pa_{\nu}\bar\phi^m\zeta^i_{\perp}\zeta^j_{\perp}.
\label{flucinducedmetric2}
\ee
Hence, the Dirac action on-shell for the expansion in Riemann normal coordinates up to second order in the fluctuations is
\be
S_{\mathrm{Dirac}} =S_{\mathrm{Dirac}}^*-\frac{M_{(p)}}{2}\int_{\phi(\Si)}\ud^{1+p} x\sqrt{|\det \bar{h}|}\left[g_{ij}(\bar\phi)\bar{h}^{\mu\nu}\pa_{\mu}\zeta^i_{\perp}\pa_{\nu}\zeta^j_{\perp}+X_{ij}\zeta^i_{\perp}\zeta^j_{\perp}\right],
\label{Diracquad}
\ee
where $S_{\mathrm{Dirac}}^*$ corresponds to the action on-shell,  $\bar{h}_{\mu\nu}= g_{ij}(\bar\phi)\pa_{\mu}\bar\phi^i\pa_{\nu}\bar\phi^j$  and
\be 
X_{ij}=-\bar{h}^{\mu\nu}R_{limj}\pa_{\mu}\bar\phi^l \pa_{\nu}\bar\phi^m-2\bar{h}^{\mu\rho}\bar{h}^{\nu\s}g_{li}(\bar\phi)g_{mj}(\bar\phi)\kappa^l_{\mu\nu}(\bar\phi)\kappa^m_{\rho\s}(\bar\phi).
\label{Xij}
\ee
Since the action is invariant under target diffeomorphisms, the quadratic part of Eq. \eqref{Diracquad}  holds for any coordinate system. From  \eqref{eom} and
\be 
\cK \equiv \bar{h}^{\mu\nu}\cK_{\mu\nu}=-\zeta^i_{\perp}g_{ij}(\bar\phi)K^j(\bar\phi),
\ee 
we see that the equations of motion imply that the mean curvature vanishes, i.e. the $p$-branes correspond to minimal submanifolds, see for example \cite{JamesSimons68,AnciauxMS}. Some remarks are in order: $i)$ the orthogonality between the vectors $\pa_\mu\bar\phi^i$ and $\zeta^i_{\perp}$ have been used instead of Dirichlet boundary conditions on the background fields, $ii)$ the extrinsic curvature terms in Eq. \eqref{Xij} vanishes for $p=0$ and $iii)$ in order to find the spectrum of the quadratic operator defined in Eq. \eqref{Diracquad},  boundary conditions on the fluctuations must still be provided. We will discuss this in the next section.

Finally, in order to study more closely  the operator of the fluctuations, let us consider the non-coordinate basis in the ambient space with the Levi-Civita spin connection $\Omega^A_{\;\;\; B}$ . This connection is uniquely determined  by the set of vectors $\{E_A=E_A^{\;\;\;i}\pa_i\}$ and their duals $\{e^A=e^A_{\;\;\; i} \ud \phi^i\}$.  
For convenience the bar on the background fields will be omitted for the remainder of this note.  The fluctuations are written as
\be
\zeta^i_{\perp}=\zeta^A_{\perp}E_A^{\;\;\; i}.
\ee
Then we continue to exploit the Riemann normal coordinate system and find 
\be
\pa_{\mu} \zeta^i_{\perp} = E_A^{\;\;\; i}\mathscr{D}_{\mu} \zeta^A_{\perp},\quad  \mathscr{D}_{\mu} \zeta^A_{\perp}=\pa_{\mu}\zeta^A_{\perp}+ \omega^A_{\;\;\; B\mu} \zeta^B_{\perp},
\ee
where $\omega^A_{\;\;\; B\mu}$ is the pullback of the spin connection, i.e. $\omega^A_{\;\;\; B\mu} =\pa_{\mu}\phi^i\Omega^A_{\;\;\; B i}$. The quadratic action becomes
\be
S^{(2)}= -\frac{M_{(p)}}{2}\int\ud^{1+p} x\sqrt{|\det h|}\left[\eta_{AB}\mathscr{D}_{\mu}\zeta^A_{\perp}\mathscr{D}^{\mu}\zeta^B_{\perp}+X_{AB}\zeta^A_{\perp}\zeta^B_{\perp}\right],
\ee
where
\be
X_{AB}= -h^{\mu\nu}q^C_{\;\;\; \mu}q^D_{\;\;\; \nu}R_{CADB}-2h^{\mu\rho}h^{\nu\s}\kappa_{A\mu\nu}(\phi)\kappa_{B\rho\s}(\phi),
\label{massmatrix}
\ee
with $q^C_{\;\;\; \mu}=\pa_{\mu}\phi^le^C_{\;\;\; l}$ and $\kappa_{A\mu\nu}(\phi)=\eta_{AB}e^B_{\;\;\; l}\kappa^l_{\mu\nu}(\phi)$. It is useful to  re-write the quadratic action as
\be
S^{(2)}= -\frac{M_{(p)}}{2}\int\ud^{1+p} x\,\d_{AB}D_{\mu}\left(\sqrt{|\det h|}\zeta^A_{\perp} D^{\mu}\zeta^B_{\perp}\right)-\frac{M_{(p)}}{2}\int\ud^{1+p} x\sqrt{|\det h|}\zeta^A_{\perp}\cO_{AB}\zeta^B_{\perp},
\label{THEQUADRATICACTION}
\ee
 where $\cO_{AB}$ is the operator of fluctuations and it is given by
 \be
 \cO_{AB}=-\d_{AB}\Box -2\omega_{AB\mu}D^{\mu}+\d_{CD} h^{\mu\nu}\omega^C_{\;\;\; A\mu}\omega^D_{\;\;\; B\nu}+X_{AB}.
 \label{THEOPERATOROFFLUC}
 \ee
In general we will assume a boundary condition so that the first term in \eqref{THEQUADRATICACTION} vanishes. In order to compute the path integral,  we set the Dirac action to be Euclidean. Therefore $S^{(2)}$ differs from its Lorentzian counterpart by a minus sign.


\section{One loop correction}
\subsection{Path integral}
We are now in a position to compute the one-loop correction to the Euclidean wavefunction  via the path integral. The integral is
\be
 \Psi \approx \ue^{-S^{*}}
 \cN \Delta_{FP}^{\mathrm{n.g.}} \int\cD\zeta_{\perp}^A\, \exp\left(-\frac{M_{(p)}}{2}\int\ud^{1+p} x\sqrt{|\det h|}\zeta^A_{\perp}\cO_{AB}\zeta^B_{\perp}\right),
 \label{1loopintegral}
\ee
where $\cN$ is the normalization constant and $\Delta_{FP}^{\mathrm{n.g.}}$ is the Fadeev-Popov determinant in the normal gauge. We have set $\hbar=1$ and thus the semiclassical limit corresponds to $M_{(p)}\to\infty$. Let use consider the following eigenvalue problem 
\be
\cO_{AB}\chi^B_n=\d_{AB}\la_n\chi^B_n. 
\label{EigenProblemO}
\ee
Boundary conditions on the (dimensionless) eigenfunctions $\chi^A_n$ have been assumed in order to obtain the real spectrum $\{\la_n\}$. The eigenfunctions are normalized as
\be
 \mu^{1+p}\int\ud^{1+p} x\sqrt{|\det h|} \d_{AB} \chi^A_n(x)\chi^B_m(x)=\| \vec{\chi}_n\|^2\d_{nm},
 \label{NormEigenFunctionsO}
\ee
where $\| \vec{\chi}_n\|$ is the norm and $\mu$ is a mass scale. Hence, we write $\zeta^A_{\perp}=\frac{1}{\mu}\sum_n c_n \chi^A_n$ and therefore the boundary condition on the eigenfunction translates to a condition on $\zeta^A_{\perp}$. This further implies the boundary conditions of the path integral. We take the measure to be 
\be 
\cD\zeta_{\perp}^A\to  \prod_{n}\phantom{}^{'}\ud c_n\,\sqrt{\frac{M_{(p)}}{2\pi \mu^{1+p}}}\| \vec{\chi}_n\|,
\ee
where the prime indicates that we have removed zero modes. Denoting $\cN'$ to be the normalization constant that absorbed $\Delta_{FP}^{\mathrm{n.g.}}$ we perform the integral and we obtain $ \Psi \approx \cN'\ue^{-S^{*}}/\sqrt{\det(\cO '/\mu^2)}$.

\subsubsection{Zero modes}
\begin{figure}[ht!]
\centering
\begin{tikzpicture}
\draw[->] (-0.1,0)--(3.5,0);
\draw[->] (0,-0.1)--(0,6);

\node at (3.8,0) {$\phi^1$};
\node at (0.1,6.3) {$\phi^2$};

\draw[->,thick] (0,0.8)--(3,0.8);
\node at (-0.3,0.8) {$\phi^2_0$};

\draw[<-,thick] (0,1.7) arc (-90:90:2cm);
\node at (-0.3,3.7) {$\phi^2_0$};
\draw[->] (0,3.7) -- (1.5,5);
\node at (.5,4.7) {$\rho_0$};
\end{tikzpicture}
\caption{Geodesics in the upper-half Poincar\'e model of hyperbolic space.}
\label{H2geo}
\end{figure}
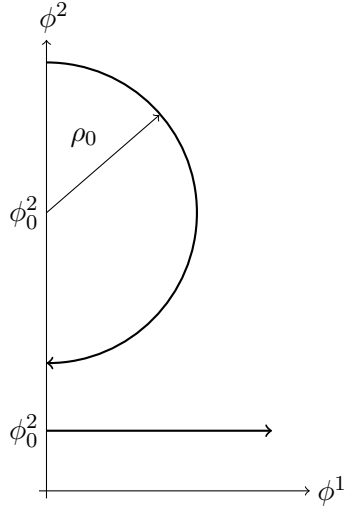

Let us assume the existence of $k$ zero modes, then  Eq. \eqref{1loopintegral} must be modified to be
\be 
 \Psi \approx \cN' \int \prod_{l=1}^k \ud c_{0}^{(l)}\,\sqrt{\frac{M_{(p)}}{2\pi \mu^{1+p}}}\| \vec{\chi}_0^{(l)}\|\,\frac{\ue^{-S^{*}}}{\sqrt{\det(\cO'/\mu^2)}}.
 \label{zero1loopintegral}
\ee
Following \cite{Gervais:1974dc,Bernard:1979qt,Dorey:2002ik,Tong:2005un}, the zero modes can be tracked from the possible family of inequivalent classical solutions. This means that the classical solution depends on $k$-independent parameters. The space of such solutions  is referred as to the moduli space of solutions. Let $\{\t^l\}$ be the set of such parameters. They are called collective coordinates and correspond to the coordinates of the moduli space. Therefore a solution is labelled as $\vec{\phi}(x,\t)$. For example, consider the geodesics in the two-dimensional Poincar\'e upper half plane as depicted in Figure \ref{H2geo}. 
From the results of section 4, the semicircle and vertical line are parametrize as
\be
\vec{\phi}_{\mathrm{semi\, c.}}=\begin{pmatrix}
\rho_0\sin(\omega x)\\
\phi_0^2+\rho_0\cos(\omega x)
\end{pmatrix},
\quad 
\vec{\phi}_{\mathrm{v.l.}}=\begin{pmatrix}
\phi^1_0\ue^{mx}\\
\phi_0^2
\end{pmatrix},
\ee
where $\omega x\in[0,\pi]$ for the former geodesic and $m x\in\mathbb{R}$ for the later. For the semicircles,  the moduli space is two dimensional: we need to specify the radius $\rho_0$ and the position of the semicircle $\phi^2_0$. Similarly for the vertical lines,   we need to specify the position of the vertical line $\phi^2_0$ and the amplitude $\phi^1_0$. 

The appearance of zero modes reflects the existence of symmetries. In our example, they reflect the invariance of the action  under some isometries of the target space (global field transformations): translational invariance along the $\phi^2$-axis is related of the position of the geodesics and the  rescaling  of the coordinates is related to the radius of the semicircle. For the vertical line, the rescaling is related to the choice of the amplitude $\phi^1_0$. Moreover, the choice of $\phi^1_0$ can equivalently interpreted as a consequence of translational invariance of the parameter space $\Sigma =\mathbb{R}$.

Let $\vec{\phi}(x,\tilde\t)$ and $\vec{\phi}(x,\t)$ be two solutions. Using the fact that the action is independent of the collective coordinates and considering $\tilde{\t}^l=\t^l+\d\t^l$, we conclude
\be
0=\frac{1}{2}\int\ud^{1+p}x \ud^{1+p}y  \left(\frac{\pa\phi^i_c}{\pa\t^l}\d\t^l\right)_x\left.\frac{\d^2 S}{\d \phi^i_x\d\phi^j_y}\right|_{\phi=\phi_c}\left(\frac{\pa\phi^j_c}{\pa\t^{l'}}\d\t^{l'}\right)_y.
\ee
The expression implies that
\be
\vec{X}^{(l)}_0=\frac{\pa\vec{\phi}_c}{\pa\t^l},
\label{zeromodesformula}
\ee
are zero modes\footnote{Since the operator  of fluctuations is the geodesic deviation operator, the zero modes are Jacobi fields (see \cite{JamesSimons68, DeWittMorette:1976up}). }. The general splitting of the field around a solution is given by
\be
\vec{\phi}(x)=\vec{\phi}(x,\t)+\vec{\zeta}_{\perp}(x,\t),
\ee
where the fluctuations depend implicitly on the collective coordinates. Since $\vec{\phi}(x)$ does not depend on the parameters,  its variation with respect $\t$ gives $\d\vec{\phi}(x,\t)=-\d\vec{\zeta}_{\perp}(x,\t)$. Notice that the zero modes $\vec{X}^{(l)}_0$ are genuine zero modes since in the ``normal'' gauge, the fluctuations and gauge transformations are orthogonal. The removal of zero modes on the mode expansion of the fluctuations can be achieved by imposing the constraints $\cC^{(l)}(\t)=0$ defined as
\bea
\cC^{(l)}(\t) \equiv \left\langle\vec{\zeta}_{\perp},\frac{\pa\vec{\phi}_c}{\pa\t^l}\right\rangle &=& \mu^{1+p}\int\ud^{1+p} x\sqrt{|\det h|} \d_{AB} \zeta^A_{\perp}(x)\chi^{(l)B}_0(x),\nn
&=&\sum_{l'} c^{l'}_0\left\langle \vec{X}^{(l')}_0 , \vec{X}^{(l)}_0\right\rangle.
\eea
The constraints  can be implemented in the path integral by inserting
\be
1=\int  \prod_{l=1}^k \ud \t^l \left|\frac{\pa \cC^{(l')}}{\pa\t^l}\right|  \prod_{l'=1}^k \d(\cC^{(l')}(\t)),
\label{trick}
\ee
where
\be
 \frac{\pa \cC^{(l')}}{\pa\t^l}=-\left\langle \vec{X}^{(l)}_0 , \vec{X}^{(l')}_0\right\rangle+\left\langle \vec{\zeta}_{\perp} ,\frac{\pa\vec{X}^{(l')}_0}{\pa\t^{l}}\right\rangle.
 \label{paCpaT}
\ee
At leading order in Eq. \eqref{paCpaT}, we obtain
\be 
 \Psi \approx \cN' \int \prod_{l=1}^k \ud \t^{l}\,\sqrt{\frac{M_{(p)}}{2\pi \mu^{1+p}}}\| \vec{\chi}_0^{(l)}\|\,\frac{\ue^{-S^{*}}}{\sqrt{\det(\cO'/\mu^2)}}.
 \label{zero1loopintegral2}
\ee
Note that the factor $\left\langle \vec{X}^{(l)}_0 , \vec{X}^{(l')}_0\right\rangle$ drops out from the determinant in Eq.\eqref{trick} due to the rescaling property of the Dirac delta function. 

For completeness of the example, the semicircle  zero modes are 
\be
\frac{\pa\vec{\phi}_c}{\pa\rho_0}=\begin{pmatrix}
\sin(\omega x)\\
\cos(\omega x)
\end{pmatrix},
\quad
\frac{\pa\vec{\phi}_c}{\pa\phi_0^2}=\begin{pmatrix}
0\\
1
\end{pmatrix}, 
\ee
and for the vertical line
\be
\frac{\pa\vec{\phi}_c}{\pa\phi_0^1}=\begin{pmatrix}
\ue^{mx}\\
0
\end{pmatrix},
\quad
\frac{\pa\vec{\phi}_c}{\pa\phi_0^2}=\begin{pmatrix}
0\\
1
\end{pmatrix}.
\ee
It is easy to see that the zero modes of the semicircle are normalizable contrary to the zero modes of the vertical line. As explained in section 4, this issue is solved by introducing a regulator $R$. Hence, we conclude that the presence of the moduli  spontaneously breaks some target isometries and the integration over the parameters recovers such invariance on the wavefunction.

\subsection{Wheeler-DeWitt equation}
We now turn to the canonical quantization approach of the system. The starting point is the Euclidean version of Eq. \eqref{Dirac}. The Lagrangian of the system is
\be 
\cL=M_{(p)}\sqrt{\det h_{\mu\nu}}=M_{(p)}\sqrt{\det h_{\hat{i}\hat{j}}}\sqrt{h_{00}-h_{0\hat{i}}h^{\hat{i}\hat{j}}h_{0\hat{j}}}
\label{DiracLagrangian}
\ee
where 
the index $\mu$ has  been split into $(0,\hat{i})$ with $\hat{i}=1,2,\ldots,p$. We choose $x^0$ to be the ``time'' coordinate. The canonical momenta are
\be
P_i=\frac{\d\cL}{\d\pa_0\phi^i}=\frac{M_{(p)}^2\det h_{\hat{i}\hat{j}}}{\cL}\left[g_{ij}(\phi)-g_{il}(\phi)\pa_{\hat{i}}\phi^lh^{\hat{i}\hat{j}}g_{jm}(\phi)\pa_{\hat{j}}\phi^m\right]\pa_0\phi^j,
\label{momentapbrane}
\ee
For $p=0$, we have $P_i=\frac{M_{(p)}^2}{\cL}g_{ij}\pa_0\phi^j$. From the momenta follows that the Hamiltonian vanishes and 
\bea
g^{ij}(\phi) P_i P_j - M_{(p)}^2\det h_{\hat{i}\hat{j}}&=&0,\label{HCgeneral}\\
P_i\pa_{\hat{i}}\phi^i &=&0 \label{RepCgeneral}.
\eea
The first expression is the $(1+p)$-dimensional generalization of a ``on-shell conditions'' of a worldline. In the context of constraint dynamics, a  vanishing Hamiltonian suggest that the system is generally covariant and it is expressed in terms of a linear combination of the constraints\footnote{A word of caution is needed. In the framework of constraint systems one needs to distinguish between ``weakly'' and ``strongly'' vanishing quantities. A ``weakly'' vanishing quantity may have non-zero Poisson brackets with the canonical variables. Therefore, by vanishing Hamiltonian we actually mean ``weakly'' vanishing. See \cite{Henneaux:1992ig}  for more details.}. This is well known  for the worldline case and also follows for a general $p$, see for example \cite{Banerjee:2005bb} and references within. The Hamiltonian density can be written as $\cH=N\cH_0+N^{\hat{i}}\cH_{\hat{i}}$ where $N, N^{\hat{i}}$ are Lagrange multipliers and 
\be
\cH_0\equiv  g^{ij}(\phi) P_i P_j - M_{(p)}^2\det h_{\hat{i}\hat{j}}, \quad \cH_{\hat{i}}\equiv P_i\pa_{\hat{i}}\phi^i,
\ee
are referred as to the Hamiltonian and momentum constraint respectively\footnote{The same constraints are obtained if we instead consider the Polyakov action. The worldvolume metric is written in an ADM form and therefore $N, N^{\hat{i}}$ corresponds to the lapse and shift functions respectively. }. As in the $p=0$ case, the Hamiltonian constraint  reflects the invariance of the action under ``time'' reparametrizations. For $p\neq 0$, the momentum constraint arises due to ``spatial''  reparametrizations of extended objects. The Hamiltonian and momentum constraint corresponds to primary constraint and we expect that they must be preserved during ``time'' evolution. In  \cite{Banerjee:2005bb} this is shown to be the case since the constraint algebra is closed, i.e. the  Poisson bracket between the constraints are proportional to linear combinations of themselves and therefore there are not secondary constraints\footnote{In the case of the Polyakov action, the primary constraints are related to the momenta of the lapse and shift functions. ``Time'' evolution gives secondary constraints which corresponds to the Hamiltonian and momentum constraint.  After  applying again the ``time'' evolution condition, no new constraint is found and the algebra of constraints is closed. Therefore the system is first class.}. Since the $1+p$ primary constraints are preserved, the system is first class and as explained in \cite{Henneaux:1992ig} these constraints generate gauge transformations, i.e. the reparametrizations. Therefore, $1+p$ gauge fixing conditions are needed to fully fix the gauge. 

Following \cite{Henneaux:1992ig},  the counting of physical degrees of freedom can be computed from 
\begin{multline}
2\times (\mathrm{Physical\, d.o.f})= \# \mathrm{Canonical\, variables}- \# \mathrm{Second\, class \,constraints}\\-\# \mathrm{First\, class\, constraints}-\#\mathrm{Gauge\, conditions}.
\end{multline}
 For the $p$-brane the number of physical degrees of freedom are $D-(1+p)$, it corresponds to the codimension of the embedded submanifold. The physical degrees of freedom are located at perpendicular directions to the submanifold. In the covariant formulation, this gives a dynamical motivation to the  ``normal'' gauge which complements the geometric perspective as discussed in Section 2.1. 
 
On the other hand, due to the non-linear nature of the constraints, the equations of motion are difficult to solve for general $p$ and background $g_{ij}$. Moreover, the classical problem requires to fix a gauge and the non-linearities imposes a challenge to find a solution. It is well known that for a Minkowski target space and $p=1$, the system is linearlized and solved in the light-cone gauge\footnote{This is the gauge in which it is known how to quantize the superstring. See \cite{brink2013principles} for more details. }.  For $p=2$, as fully discussed in \cite{Hoppe_2012} and reference within, non-linearities are only simplified in the light-cone gauge and the  2-brane can be approximated into a matrix model. 

 In our case, it will be shown that  these classical issues are  circumvented by considering a  limit in which  the $(1+p)$-dimensional brane effectively becomes tensionless, i.e. we will focus on the high energy limit. Roughly, the Hamiltonian constraint will become $g^{ij}P_i P_j$. Since in  Poincar\'e coordinates the metric $g_{ij}$ is conformally flat,  the Hamiltonian and momentum constraints are reduced to the ones in a flat background. Classical tensionless $(1+p)$-dimensional branes has been discussed in the literature, see for example \cite{Schild-1977,Karlhede-1986,Amorim:1987bk,BarcelosNeto:1989gs,J-1990,Hassani:1994rf}, and for $p=1$ the so called Schild's gauge,  $N=\mathrm{const.}$ and $N^{1}=0$, corresponds physically to an infinite set of free massless relativistic particles streaming orthogonally to the 1-brane. The generalization of this gauge for $p\geq 2$ is straightforward and due to its physical interpretation, it should be equivalent to the ``normal'' gauge in the covariant description after identifying the normal fluctuations as massless relativistic particles in the high energy limit.

 For the quantization, we adopt the Schr\"odinger picture and follow \cite{Symanzik:1981wd,Luscher:1985iu,Jackiw:1988sf}.  Then, if $\Sigma=\mathbb{R}^{1+p}$, the wave functional $\Psi[\varphi^i(x^{\hat{i}})]$ has the interpretation that $|\Psi[\varphi^i(x^{\hat{i}})]|^2$ is proportional to the probability for the quantum fields $\hat{\phi}^i(x^0,x^{\hat{i}})$ to assume the classical values $\varphi^i(x^{\hat{i}})$ at ``time'' $x^0=0$.  This implies
\be
 \hat{\phi}^i(0,x^{\hat{i}})\Psi[\varphi(x^{\hat{i}})]=\varphi^i(x^{\hat{i}})\Psi[\varphi^i(x^{\hat{i}})].
 \label{ScHfieldtheory}
\ee
In a general space $\Sigma$, we follow the same reasoning.  As stated in \cite{Jacobson:2003vx}, the canonical commutation relation at equal time is
\be
[ \hat{\phi}^i(0,x^{\hat{i}}), \hat{P}_j(0,y^{\hat{i}})]=i\hbar\,\d^i_j\d(x^{\hat{i}},y^{\hat{i}}), 
\ee 
where the Dirac delta  is a density of weight one in the second argument and is defined, without the use of a metric, as  $\int\ud^{d-1}y\,  \d(x^{\hat{i}},y^{\hat{i}}) f(y^{\hat{i}})=f(x^{\hat{i}})$ for any scalar function $f$. The canonical commutation relation and Eq. \eqref{ScHfieldtheory} imply
\be
 \hat{P}_i=\frac{\hbar}{i}\frac{\d}{\d\varphi^i}.
\ee

We are now in a position to quantize Eq. \eqref{HCgeneral} and Eq. \eqref{RepCgeneral}. The first issue we need to address is an ordering problem in these expressions. We choose an ordering in which the momenta acts first on the wavefunctional, i.e. 
\bea
 \left[g^{ij}(\varphi)\frac{\d^2}{\d \varphi^i \d \varphi^j}+\frac{M^2_{(p)}}{\hbar^2}\det h_{\hat{i}\hat{j}}\right]\Psi[\varphi^i]&=&0, \label{FULLWdWeq}\\
\pa_{\hat{i}}\phi^i \frac{\d}{\d\varphi^i} \Psi[\varphi^i] &=&0 \label{RepInvWF}.
\eea
The first line corresponds to the Euclidean  Wheeler-DeWitt equation and the second line indicates that the wavefunctional is reparametrization invariant under the ``spatial'' coordinates. For $p=0$, we replace $M^2_{(p)}\det h_{\hat{i}\hat{j}}\to M^2_{(0)}$ in the Wheeler-DeWitt equation. For $p\neq 0$, this equation faces the issue of singularities because the expression
\be
 \frac{\d^2\Psi[\varphi^i]}{\d \varphi^i (x^{\hat{i}})\d \varphi^j(y^{\hat{i}})},
\ee
becomes singular as $y\to x$. Therefore, we introduce an ultraviolet regulator so that the Wheeler-DeWitt equation is well defined.  We address this issue as in  \cite{Mansfield:1993pd,1996NCimB.111...85H} where the functional derivative is regulated as 
\be
  \frac{\d^2}{\d \varphi^i (x^{\hat{i}})\d \varphi^j(y^{\hat{i}})}\to \int\ud^py\, K(x^{\hat{i}},y^{\hat{i}},\tau)\frac{\d^2}{\d \varphi^i (x^{\hat{i}})\d \varphi^j(y^{\hat{i}})},
\ee
where the heat kernel $K(x^{\hat{i}},y^{\hat{i}},\tau)$, a bi-tensor of density weight one, satisfies some heat equation with the initial condition
\be
\lim_{\tau\to 0}K(x^{\hat{i}},y^{\hat{i}},\tau)=\d (x^{\hat{i}},y^{\hat{i}}).
\ee 
For the moment we do not specify the equation for $K$ but assume the existence of its solution. Moreover, we work with the associated (regulated) Green function defined as, see \cite{grigoryan2009heat}, 
\be
G_{\Lambda} (x^{\hat{i}},y^{\hat{i}}) =\int\limits_0^{\infty} \ud\tau\, \ue^{-\Lambda^2\tau}K(x^{\hat{i}},y^{\hat{i}},\tau),
\ee
where $\Lambda$ is a mass scale. Setting $\Psi=\ue^{-\cS/\hbar}$, Eq. \eqref{FULLWdWeq} becomes 
\begin{multline}
 \int\ud^py\, G_{\Lambda}(x^{\hat{i}},y^{\hat{i}})g^{ij}(\varphi(x^{\hat{i}}))\left[\frac{\d\cS}{\d \varphi^i (x^{\hat{i}})}\frac{\d\cS}{\d \varphi^j(y^{\hat{i}})}-\hbar \frac{\d^2\cS}{\d \varphi^i (x^{\hat{i}})\d \varphi^j(y^{\hat{i}})}\right]\\+\frac{M^2_{(p)}}{\Lambda^2}\det h_{\hat{i}\hat{j}}(x^{\hat{k}})=0.
\end{multline}
As stated in section 1, in the limit $\hbar\to 0$  we obtain the regulated Euclidean Hamilton-Jacobi equation. For the one-loop correction via the path integral, we set $\hbar=1$ and take $M_{(p)}\to\infty$. Thus, the expansion in powers of $\hbar$ of $\cS$ is replaced by
\be
\cS=M_{(p)}\cW_0+\cW_1+\frac{1}{M_{(p)}}\cW_2+\ldots. 
\ee
A more convenient parametrization is $\Psi = A \ue^{-M_{(p)}\cW_0-\frac{1}{M_{(p)}}\cW_2-\ldots}$ where $ \cW_1=-\ln A$. The equations up to one-loop are
\bea
\int\ud^py\, G_{\Lambda}(x^{\hat{i}},y^{\hat{i}})g^{ij}(\varphi(x^{\hat{i}}))\frac{\d\cW_0}{\d \varphi^i (x^{\hat{i}})}\frac{\d\cW_0}{\d \varphi^j(y^{\hat{i}})}+\frac{1}{\Lambda^2}\det h_{\hat{i}\hat{j}}(x^{\hat{k}})&=&0,\nn
  \int\ud^py\, G_{\Lambda}(x^{\hat{i}},y^{\hat{i}})g^{ij}(\varphi(x^{\hat{i}}))\frac{1}{A}\left[2\frac{\d\cW_0}{\d \varphi^{(i} (x^{\hat{i}})}\frac{\d A}{\d \varphi^{j)}(y^{\hat{i}})}-\frac{\d^2\cW_0}{\d \varphi^i (x^{\hat{i}})\d \varphi^j(y^{\hat{i}})}A\right]&=&0.
\eea
We have assumed that  
\be
\frac{1}{A}\frac{\d^2 A}{\d\varphi ^i\d\varphi ^j}\approx 0, 
\ee
i.e. $A$ varies slowly. Therefore, the WKB approximation is valid under these circumstances. This solution must satisfy Eq. \eqref{RepInvWF} as well and this implies that $\cS$ must be invariant under ``spatial'' reparametrizations. 

The propagator inherits from the heat kernel a high energy expansion of the form
\be
 G_{\Lambda}(x^{\hat{i}},y^{\hat{i}}) = \d(x^{\hat{i}},y^{\hat{i}})+O\left(\frac{1}{\Lambda}\right),
\ee 
and we assume the following expansion for the functions
\be 
\cW_l=\cW_l^{\infty}+O\left(\frac{1}{\Lambda}\right), \quad l=1,2,\ldots.
\ee
Hence, at leading order, the equations become
\bea
 g^{ij}(\varphi)\frac{\d\cW_0^{\infty}}{\d \varphi^i }\frac{\d\cW_0^{\infty}}{\d \varphi^j}&=&0, \label{pdimeq}\\
 2g^{ij}\frac{\d\cW^{\infty}_0}{\d \varphi^i }\frac{\d A^{\infty}}{\d \varphi^j}-g^{ij}\frac{\d^2\cW^{\infty}_0}{\d \varphi^i\d \varphi^j}A^{\infty}&=&0 \label{oneloopeqdeterminant}.
\eea
In order to avoid a possible order of limits confusion, we stress that the above equations have been computed for finite $M_{(p)}$ and $\Lambda\to \infty$. The semiclassical limit will be taken after these equations are solved. Notice that Eq. \eqref{pdimeq} corresponds to the Hamiltonian constraint of a tensionless $(1+p)$-dimensional brane since $\frac{\d\cW^{\infty}_0}{\d \varphi^i}$ is proportional to the classical momentum $P_i$ in the Hamilton-Jacobi formalism. From reparametrization invariance and dimensional analysis, the ansatz for $\cW_0^{\infty}$ is taken to be
\be 
\cW_0^{\infty}=\frac{\a}{\mu}\int\ud^p x\sqrt{\det h_{\hat{i}\hat{j}}(x^{\hat{k}})},
\label{solofpdimeq}
\ee
where $\a$ is a dimensionless constant and $\mu$ is a mass scale. This is a solution of Eq. \eqref{pdimeq} if $\varphi^i$ describe a $p$-dimensional minimal surfaces. We expect that this solution can be constructed from the $(p+1)$-dimensional minimal surfaces since if we restrict the Lagrangian  given in Eq. \eqref{DiracLagrangian} at $x^0=0$ and assume that 
\be 
\left.\sqrt{h_{00}-h_{0\hat{i}}h^{\hat{i}\hat{j}}h_{0\hat{j}}}\right |_{x^0=0}=\mathrm{constant},
\ee
we  recover the Lagrangian in Eq. \eqref{solofpdimeq}. On the other hand, evaluating at $x^0=0$ the equations of the $(p+1)$-dimensional minimal surface given in Eq. \eqref{eom}, we obtain the equations derived from Eq. \eqref{solofpdimeq} if we assume 
\bea
\left.\left(\pa_0(\sqrt{\det h_{\mu\nu}}h^{00}\pa_0\phi^i)+\pa_0(\sqrt{\det h_{\mu\nu}}h^{0\hat{i}}\pa_{\hat{i}}\phi^i)+\pa_{\hat{i}}(\sqrt{\det h_{\mu\nu}}h^{\hat{i}0}\pa_0\phi ^i)\right)\right |_{x^0}&=&0,\nn
&&\\
\left.\left(h^{00}\pa_0\phi^l\pa_0\phi^m+2h^{0\hat{i}}\pa_{\hat{i}}\phi^l\pa_0\phi^m\right)\right |_{x^0}&=&0.\nn
&&
\eea
This can be satisfied if we set $h^{0\hat{i}}=0$ and $h^{00}=0$ at $x^0=0$. Writing the worldvolume metric in an ADM form, we see that $h^{0\hat{i}}=0$ is achieved for all points if we choose $N^{\hat{i}}=0$. The second condition $h^{00}=0$ cannot be extended to all points and more important it sets the difference between a flat and hyperbolic background. Moreover, it also sets the difference between Schild's gauge. In the hyperbolic background the lapse function is not a constant since is proportional to the conformal factor of the background metric in Poincar\'e coordinates. Nevertheless, at the boundary  we do recover the same geometrical set up of Schild's gauge. The physical interpretation of the orthogonal streaming of massless particle interpretation holds beyond the boundary. 

To illustrate this solution, we focus on the $p=1$ case and a 3-dimensional hyperbolic background. It is known that semi-spheres are minimal surfaces and in terms of $\phi^i(x^0,x^1)$ they can be parametrized as
\bea
\phi^1(x^0,x^1)&=&\rho_0\sin(\omega x^0),\nn
\phi^2(x^0,x^1)&=&\rho_0\cos(\omega x^1)\cos(\omega x^0)+\phi^2_0,\nn
\phi^3(x^0,x^1)&=&\rho_0\sin(\omega x^1)\cos(\omega x^0)+\phi^3_0, 
\label{semisphereparametrization1}
\eea  
with $\omega x^1\in [0,2\pi]$ and $\omega x^0\in [0,\pi]$. Then for $x^0=0$ we find
\bea
\varphi^1(x^1)&=&0,\nn
\varphi^2(x^1)&=&\rho_0\cos(\omega x^1)+\phi^2_0,\nn
\varphi^3(x^1)&=&\rho_0\sin(\omega x^1)+\phi^3_0,
\label{pdimsolutionsCWL}
\eea  
and 
\be
\left. h_{00}\right |_{x^0=0}=\left. \frac{\rho_0^2\omega^2L^2}{(\phi^1)^2}\right |_{x^0=0}=\frac{\rho_0^2\omega^2L^2}{\epsilon^2}, \quad \left. h_{0\hat{i}}\right |_{x^0=0}=0,
\ee
where $\epsilon$ is a cut-off for $\varphi^1=0$. Then the solutions given by Eq. \eqref{pdimsolutionsCWL} gives
\be
\left.  \cW_0^{\infty}\right|_{\mathrm{on-shell}}=\frac{\a}{\mu}\left(\frac{L}{\epsilon}\right)\times(\mathrm{Perimeter\,of\, the \, circle}).
\label{perimeterlaw}
\ee
 For a general $p$, we argue that Eq. \eqref{solofpdimeq} is a solution in a hyperbolic background for $\varphi^1\to 0$. In order to  support this claim,  we study the one-loop correction.


Note that Eq. \eqref{oneloopeqdeterminant} can be written as 
\be
 g^{ij}\frac{\d}{\d \varphi^i}\left(\frac{\d\cW^{\infty}_0}{\d \varphi^j}(A^{\infty})^2\right)=0.
\ee
This implies 
\be
 \mu^p\frac{\d\cW^{\infty}_0}{\d \varphi^i}(A^{\infty})^2= n_i,
 \label{oneloopsolWKB}
\ee
where $n_i$ is a dimensionless constant and finite vector. Since 
\bea 
\frac{\d\cW^{\infty}_0}{\d \varphi^i}&=&\frac{\a}{\mu}\int\ud^p x\sqrt{\det h_{\hat{i}\hat{j}}}g_{ij}(\varphi)E^j,
\label{dW0dphi}
\eea
where $E^j$ is the $p$-dimensional analogue of Eq.\eqref{eom}, Eq. \eqref{pdimeq} implies that the amplitude $A^{\infty}$ must diverge. In this limit, the WKB approximation ceases to be valid\footnote{This is the functional analogue of being close to the turning points. In the next section, we discuss this for the worldline in detail.}. A reparametrization invariant ansatz of $A^{\infty}$ that satisfies the on-shell divergence behaviour is $A^{\infty}=(\mu^p \cW^{\infty}_0  )^{\frac{a}{2}}$ for a constant $a$. Recall that $ \left.  \cW_0^{\infty}\right|_{\mathrm{on-shell}}\to \infty$ as $\varphi^1\to 0$. Then  Eq. \eqref{oneloopsolWKB} result
\be
 \frac{\d}{\d \varphi^i}\left((\mu^p\cW^{\infty}_0)^{a+1}\right)= n_i(a+1).
\ee
The choice $a=-1$ is a solution for any $n_i$. Thus  we have shown that taking $\varphi^1\to 0$ in  Eq. \eqref{solofpdimeq} for general $p$ is necessary and as discussed for $p=1$ in Eq. \eqref{perimeterlaw} we introduce a cutoff $\epsilon$.

Hence, we conclude that the WKB approximation is valid in a hyperbolic background at $\varphi^1\to 0$. Moreover, at leading order, the regulated functions $\cW(\epsilon)$ have the form
\bea
\left.  \cW_0^{\infty}\right|_{\mathrm{on-shell}}&\approx &\frac{\a}{\mu}\left(\frac{L}{\epsilon}\right)^p\times(p\mathrm{-dimensional\, volume})+\mathrm{finite\, part},\label{HRclass}\\
 \left.  \cW_1^{\infty}\right|_{\mathrm{on-shell}}&\approx & \frac{1}{2}\ln\left(\mu^{p}\left(\frac{L}{\epsilon}\right)^p\times (p\mathrm{-dimensional\, volume})\right)+\mathrm{finite\, part}\label{HRoneloop}.
\eea
As $\epsilon\to 0$ we obtain $\cS_1\to \infty$ and we conclude that at leading order there are no $\cS_1\to -\infty$ divergences near the boundary. Notice that the classical divergence of the string in AdS has already computed in the holographic renormalization framework \cite{Papadimitriou:2010as}. Therefore, the semiclassical Wheeler-DeWitt solution extends the result to one-loop, i.e. the solution indicates a logarithmic divergence at one-loop that depends on the $p$-dimensional volume and that does not break the semiclassical approximation.

\section{Quantum Mechanics: Worldline in $d=2$}
With the purpose of applying the theory developed in the previous sections, we study in full detail the spectrum of the operator associated with the fluctuations for the simplest but non-trivial case: a worldline in a two-dimensional ambient space. The metric is written as
\be
g=f^2(\phi^1,\phi^2)\left[(\ud \phi^1)^2+(\ud \phi^2)^2\right]. 
\label{metricd2}
\ee
The operator of fluctutations is given by
\be
\cO =-\frac{\ud^2}{\ud x^2}\mathbb{I}-2iF\s_2\frac{\ud}{\ud x} +W\s_1+F^2\mathbb{I}+\mathbb{V},
\label{formofOind2}
\ee
where 
\be
\s_1 =\begin{pmatrix}
0&1\\
1&0
\end{pmatrix},\,
\s_2 =\begin{pmatrix}
0&-i\\
i&0
\end{pmatrix},\,
\mathbb{V}
=\begin{pmatrix}
(\dot\phi^2)^2\Box \ln f&0\\
0&(\dot\phi^1)^2\Box \ln f
\end{pmatrix},\,
W=-\dot\phi^1\dot\phi^2\Box \ln f,
\ee
and $\dot\phi^1 \equiv\frac{\ud \phi^1}{\ud x}$ and $\omega^1_{\;\;\;2x}\equiv F=\dot{\phi}^1\pa_2\ln f -\dot{\phi}^2\pa_1\ln f$. The details of this result are shown in Appendix \ref{App1}. 
In order to solve the eigenvalue problem $\cO_{AB} \chi^B_n =\la_n\d_{AB} \chi^B_n$ we follow \cite{LevSa91} (chapter 7) and consider the $2\times 2$ matrix $P(x)$ and the vector $\vec{u}_{n}=P^{-1}\vec{\chi}_{n}$. Assuming that $P$ is of the form
\be
P(x)=\ue^{-i\s_2G(x)} =\begin{pmatrix}
\cos G(x)&-\sin G(x)\\
\sin G(x)&\cos G(x)
\end{pmatrix}, \quad G(x)=G(x_0)+\int\limits_{x_0}^x\ud x'\, F(x'), 
\ee
the resulting eigenvalue problem for $\vec{u}_n$ is
\be
-\frac{\ud^2\vec{u}_{n}}{\ud x^2}+\mathbb{U}\vec{u}_{n}=\la_n\vec{u}_{n},\quad \mathbb{U}=\ue^{i\s_2G(x)}\mathbb{M}\ue^{-i\s_2G(x)},
\label{eigenprobforu}
\ee
where $\mathbb{M}= W\s_1+i\dot{F}\s_2+\mathbb{V}$. The matrix $\mathbb{U}$ become diagonal 
\be
\mathbb{U}=\Box\ln f\begin{pmatrix}
(\dot{\phi}^2\cos G-\dot\phi^1\sin G)^2&0\\
0&(\dot{\phi}^1\cos G+\dot\phi^2\sin G)^2
\end{pmatrix} ,
\ee
if and only if the geodesics satisfy
\be
\frac{\dot\phi^1\dot\phi^2}{(\dot\phi^1)^2-(\dot\phi^2)^2}=\frac{1}{2}\tan(2G). 
\label{thediagcon}
\ee
Explicit constructions of the matrix $\mathbb{U}$ are given in the following examples.
\subsection{Example 1}
Let us set $\phi^2$ to a constant. The geodesic equations are
\be
 \ddot{\phi}^1=-\pa_1\ln f (\dot{\phi}^1)^2,\quad \pa_2\ln f (\dot{\phi}^1)^2=0.
\ee
The non-trivial solution requires $\pa_2 f=0$. For this choice, we obtain $F=0$ and
\be
\int\ud \phi^1\, f(\phi^1)=C_1x+C_2.
\ee
The diagonalization condition given in Eq. \eqref{thediagcon} is satisfied for
\be 
G(x_0)=0,\pm \frac{\pi}{2},\pm \pi,\pm \frac{3\pi}{2},\ldots.
\ee
Similar results are obtained by setting $\phi^1$ to a constant. Note that the diagonalization is achieved due to the presence of a continuous isometry generated by the Killing vector  $K_2=\pa/\pa\phi^2$  for $f=f(\phi^1)$ and  a Killing vector  $K_1=\pa/\pa\phi^1$ for $f=f(\phi^2)$. The resulting matrix $\mathbb{U}$ for $f=f(\phi^1)$ is
\be
\mathbb{U}=\begin{pmatrix}
0&0\\
0&(\dot\phi^1)^2 \Box\ln f
\end{pmatrix} .
\ee
\subsection{Example 2}
We can study other types of solutions assuming translational invariance along one of the axis. For the choice $f=f(\phi^1)$, the general geodesic equations are
\bea 
\ddot{\phi}^1&=& \pa_1\ln f \left[(\dot{\phi}^2)^2-(\dot{\phi}^1)^2\right],\label{thegeqiso1}\\
\ddot{\phi}^2&=& -2\pa_1\ln f \dot{\phi}^1\dot{\phi}^2. \label{thegeqiso2}
\eea
For $f=f(\phi^2)$, we interchange $1\leftrightarrow 2$ in Eq. \eqref{thegeqiso1} and Eq. \eqref{thegeqiso2}. The equations can be rewritten as 
\bea
\frac{\ud}{\ud x}\left[f\dot{\phi}^1-\pa_1 f\phi^2\dot{\phi}^2\right]&=&f\left(\ddot{\phi}^1-\pa_1\ln f \left[(\dot{\phi}^2)^2-(\dot{\phi}^1)^2\right]\right)\nn
&&-\phi^2\dot\phi^1\dot\phi^2\left(\pa_1^2 f-2\frac{(\pa_1 f)^2}{f}\right),\label{total1}\\
\frac{\ud}{\ud x}\ln \left(\dot\phi^2 f^2\right)&=&0. 
\eea
Assuming,
\be
 \pa_1^2 f-2\frac{(\pa_1 f)^2}{f}=0,
 \label{thefcondition}
\ee
the equations can be simplified to 
\be
f\dot{\phi}^1-\pa_1 f\phi^2\dot{\phi}^2=D_1,\quad \dot\phi^2=\frac{D_2}{f^2}, 
\label{theequationsintegral}
\ee
where $D_1$ and $D_2$ are constants. It remains to find the possible solutions of the condition given in Eq. \eqref{thefcondition}. Let $z=\pa_1\ln f$, then the condition becomes the Bernoulli equation $\pa_1 z=z^2$. After solving this equation we obtain
\be
f(\phi^1)=\frac{B_2}{B_1-\phi^1}. 
\ee
Taking the ratio of the equations in Eq. \eqref{theequationsintegral}, we conclude that the geodesics are of the form
\be
(\phi^1-B_1)^2+(\phi^2-\phi^2_0)^2=\rho_0^2, 
\ee
where $\phi^2_0$ and $\rho_0$ are constants. We have two equivalent parametrizations of the fields
\be
\mathrm{I}:\quad \phi^1=B_1+\rho_0\cos(\omega x),\quad  \phi^2=\phi^2_0+\rho_0\sin(\omega x),
\ee
and
\be
\mathrm{II}:\quad \phi^1=B_1+\rho_0\sin(\omega x),\quad  \phi^2=\phi^2_0+\rho_0\cos(\omega x),
\ee
with $\omega>0$. Direct substitution shows that $F=\omega$ and $F=-\omega$ for the parametrization $\mathrm{I}$ and $\mathrm{II}$ respectively. In order to satisfy Eq. \eqref{thediagcon} we must choose $G(x_0)=\omega x_0+k\pi/2$ and $G(x_0)=-\omega x_0+k\pi/2$ where $k\in\mathbb{Z}$ for  $\mathrm{I}$ and $\mathrm{II}$ respectively. The resulting $\mathbb{U}$ matrices are
\be
 \mathbb{U}_{\mathrm{I}}=\begin{pmatrix}
 \frac{\omega^2}{\cos^2(\omega x)}&0\\
 0&0
 \end{pmatrix},
\quad
 \mathbb{U}_{\mathrm{II}}=\begin{pmatrix}
 0&0\\
 0&\frac{\omega^2}{\sin^2(\omega x)}
 \end{pmatrix}.
\ee

\subsection{A more general strategy}
From Eq. \eqref{eigenprobforu} we learned that  it is required to solve the generic Schr\"odinger equation
\be
\ddot{\Psi}+(\la-V)\Psi=0. 
\ee
Following \cite{Levai:1989eaa}, we set $\Psi(x)=\cA(x)\cF(\psi(x))$ to study the structure of solutions. The equation becomes 
\be
\cF''+Q(\psi(x))\cF'+R(\psi(x))\cF=0, 
\label{eqofF}
\ee
where $\cF'=\frac{\pa\cF}{\pa \psi}$ and
\be
 Q(\psi(x))=\frac{2}{\dot\psi}\frac{\dot\cA}{\cA}+\frac{\ddot\psi}{\dot\psi^2},\quad R(\psi(x))=\frac{1}{\dot\psi^2}\frac{\ddot\cA}{\cA}+\frac{\la-V}{\dot\psi^2},
 \label{eqforQR}
\ee
or
\be
 \la-V=\frac{1}{2}\{\psi,x\}+\dot\psi^2\left[R(\psi)-\frac{1}{2}\frac{\pa Q}{\pa \psi}-\frac{1}{4}Q^2\right],\quad 
 \cA(x)=\cA(x_0)\sqrt{\frac{\dot\psi(x_0)}{\dot\psi(x)}}\exp\left(\frac{1}{2}\int\limits_{\psi(x_0)}^{\psi(x)}\ud\tilde{\psi}\,Q(\tilde{\psi})\right),
\ee
where $\{\psi,x\}$ is the Schwarzian derivative. The key property of the parametrization is that  Eq.  \eqref{eqofF} can be solved analytically if $\cF$ corresponds to a special function. As developed in \cite{Levai:1989eaa}, we consider
\be
 Q(\psi)=\frac{\b-\a-(\a+\b+2)\psi}{1-\psi^2}, \quad R(\psi)=\frac{n(n+\a+\b+1)}{1-\psi^2}.
 \label{defofJP}
\ee
The solution  of \eqref{eqofF} is of the form
\be
\Psi^{(\a,\b)}_n(x)= \cA(x_0) \sqrt{\frac{\dot\psi(x_0)}{\dot\psi(x)}}\left(\frac{1-\psi(x)}{1-\psi(x_0)}\right)^{\frac{\a+1}{2}}\left(\frac{1+\psi(x)}{1+\psi(x_0)}\right)^{\frac{\b+1}{2}} P_n^{(\a,\b)}(\psi(x)),
\ee
where  $\psi(x)\in [-1,1]$ and $P_n^{(\a,\b)}(\psi(x))$ are the  Jacobi polynomials with $\a,\b>-1$. In order to see which potential $V$ we can obtain, the specific form of $\psi(x)$ must be given. On the other hand, the relation of this method with supersymmetric quantum mechanics is realized by assuming $\dot\psi^2R(\psi)=C=\mathrm{constant}$  and defining the superpotential as
\be
W=-\frac{\dot\cA}{\cA}. 
\label{supP}
\ee
Then 
\be
\la-V=-W^2+\dot{W}+C. 
\ee
This expression gives the supersymmetric potential $V_-$, see \cite{doi:10.1142/4687}. The combination of both approaches give us a powerful tool to compute the solution.  The condition $\dot\psi^2R(\psi)=C$ implies that
\be
 \frac{\dot\psi^2}{1-\psi^2}=c=\mathrm{constant},\quad \dot\psi^2R(\psi)=cn(n+\a+\b+1).
 \label{condition1and2}
\ee
Using Eq. \eqref{eqforQR}, \eqref{defofJP} and \eqref{supP} we get
\be
\ddot{\psi}-2W\dot\psi -c(\b-\a-(\a+\b+2)\psi) =0
\label{eqbothmethods}
\ee
Therefore, the procedure to obtain a solution of the Schr\"odinger equation is to first define the superpotential  $W$ that gives $V_-$ and then solve Eq \eqref{eqbothmethods}.  It remains to show that the solution is normalizable. Let us consider
\begin{multline}
\int\ud x\,  \Psi^{(\a,\b)}_n(x) \Psi^{(\a,\b)}_m(x)= \cA^2(x_0) \dot\psi(x_0)\left(\frac{1}{1-\psi(x_0)}\right)^{\a+1}\left(\frac{1}{1+\psi(x_0)}\right)^{\b+1}\times\\
 \int\ud x \frac{1}{\dot\psi(x)} (1-\psi^2(x)) (1+\psi(x))^{\a}(1-\psi(x))^{\b}P_n^{(\a,\b)}(\psi(x))P_m^{(\a,\b)}(\psi(x)).
\end{multline}
Using the first expression in Eq. \eqref{condition1and2} the integral becomes
\begin{multline}
\int\ud x\,  \Psi^{(\a,\b)}_n(x) \Psi^{(\a,\b)}_m(x)= \cA^2(x_0) \dot\psi(x_0)\left(\frac{1}{1-\psi(x_0)}\right)^{\a+1}\left(\frac{1}{1+\psi(x_0)}\right)^{\b+1}\times\\
\frac{1}{c} \int\ud \psi\,  (1+\psi(x))^{\a}(1-\psi(x))^{\b}P_n^{(\a,\b)}(\psi(x))P_m^{(\a,\b)}(\psi(x)).
\end{multline}
The remaining integral is known and we can choose $\cA(x_0)$ to the obtain an orthonormal basis. Hence, the full solution is written as
\be
\Psi(x)=\sum_n a_n \Psi^{\a,\b}(x). 
\ee
As an example, let us consider $W=-B\cot(\omega x)$ where $B$ is independent of $x$. The potential is
\be
\la -V=C+B^2-\frac{B(B-\omega)}{\sin^2(\omega x)}. 
\ee
If $B(B-\omega)=\omega^2$, $c=\omega^2$, $\a=\b$ and $\a+1=\frac{1}{2}-\frac{B}{\omega}$, the solution of Eq. \eqref{eqbothmethods} is $\psi(x)=\cos(\omega x)$ with $0\leq \omega x \leq \pi$. The eigenvalues are $\la= n(n-2B(\omega)/\omega)\omega^2+B^2(\omega)$. In order to use the Jacobi polynomials, we are required to impose $\a>-1$. Since $B_{\pm}=(\omega/2)(1\pm\sqrt{5})$, the constraint becomes $-(1/2)(1\pm\sqrt{5})>-1$. Thus we choose $B_-=(\omega/2)(1-\sqrt{5})$ and the solution is of the form
\be
\Psi(x)=\sum_{n=0}^{\infty}a_n \left[\sin(\omega x)\right]^{\frac{1}{4}(\sqrt{5}-1)}P^{(\frac{1}{2}(\sqrt{5}-1),\frac{1}{2}(\sqrt{5}-1))}_n(\cos(\omega x)),
\label{sinwavefunc}
\ee
with $0\leq \omega x \leq \pi$. The wavefunction $\Psi$ vanishes at $x=0$ and $x=\pi/\omega$. 
%
After establishing tools and concrete formulae from the examples, we study a concrete example. 

\subsection{The Poincar\'e half plane} 
\label{PHPsection}
The geodesics of the two-dimensional upper-half Poincar\'e model of hyperbolic space are shown in Figure \ref{H2geo}. The metric of this space is given by
\be
\ud s^2 =\frac{L^2}{(\phi^1)^2}\left[(\ud \phi^1)^2+(\ud \phi^2)^2\right], 
\ee
where $L$ corresponds to the radius.
The semicircles are already studied in Example 2. To obtain the solution, we must set $B_1=0$ and $B_2=-L$. Using the parametrization $\mathrm{II}$ of the aforementioned example, the associated spectral problem is
\be
\ddot{u}^1_{n}+\la_n u^1_{n}=0,\quad  \ddot{u}^2_{n}+\left(\la_n-\frac{\omega^2}{\sin^2(\omega x)}\right) u^2_{n}=0.
\label{UeqSemCir}
\ee
The solutions are 
\bea
 u^1_n(x)&=&a_n \cos(\sqrt{\la_n}x)+b_n \sin(\sqrt{\la_n}x),\\
 u^2_n(x)&=&c_n \left[\sin(\omega x)\right]^{\frac{1}{4}(\sqrt{5}-1)}\cC^{\left(\frac{\sqrt{5}}{2}\right)}_n(\cos(\omega x)),\\ 
 \la_n &=&n(n-1+\sqrt{5})\omega^2+\frac{\omega^2}{4}(1-\sqrt{5})^2,
\eea
with $\omega>0$, $0\leq \omega x\leq \pi$ and $C^{\left(\frac{\sqrt{5}}{2}\right)}_n(\cos(\omega x))$ are the Gegenbauer polynomials\footnote{The relation of the Gegenbauer polynomials with the Jacobi polynomials, see \cite{NIST:DLMF}, is
\be
C_{n}^{{(\alpha )}}(x)={\frac  {(2\alpha )_{n}}{(\alpha +{\frac  {1}{2}})_{{n}}}}P_{n}^{{(\alpha -1/2,\alpha -1/2)}}(x),
\ee
where $(\a)_{n}={\frac {\Gamma (\a+n)}{\Gamma (\a)}}$. 
}
Hence, the fluctuations are
\bea
 \zeta_{\perp}^1(x)&=&\sum_{n=0}^{\infty}\left[a_n \cos(\sqrt{\la_n}x)+b_n \sin(\sqrt{\la_n}x)\right]\cos(\omega x)\nn
 &&+\sum_{n=0}^{\infty}c_n \sin(\omega x)\left[\sin(\omega x)\right]^{\frac{1}{4}(\sqrt{5}-1)}C^{\left(\frac{\sqrt{5}}{2}\right)}_n(\cos(\omega x)),\\
 \zeta_{\perp}^2(x)&=&-\sum_{n=0}^{\infty}\left[a_n \cos(\sqrt{\la_n}x)+b_n \sin(\sqrt{\la_n}x)\right]\sin(\omega x)\nn
 &&+\sum_{n=0}^{\infty}c_n \cos(\omega x)\left[\sin(\omega x)\right]^{\frac{1}{4}(\sqrt{5}-1)}C^{\left(\frac{\sqrt{5}}{2}\right)}_n(\cos(\omega x)).
\eea
We set $a_n=b_n=0$ so that the fluctuations $\vec{\zeta}_{\perp}(x)$ vanish at $x=0$ and $x=\pi/\omega$ and we are left with
\be
 \vec{\zeta}_{\perp} =
\begin{pmatrix}
\sin(\omega x)\\
\cos(\omega x)
\end{pmatrix}
\left[\sin(\omega x)\right]^{\frac{1}{4}(\sqrt{5}-1)} \sum_{n=0}^{\infty}c_n C^{\left(\frac{\sqrt{5}}{2}\right)}_n(\cos(\omega x))
\ee
The coefficients $c_n$ are restricted by demanding convergence of the above sum and from $\| \vec{\zeta}_{\perp} \| \ll1$. For the second restriction, we use the definition of the Gegenbauer polynomials 
\be
 C^{\left(\frac{\sqrt{5}}{2}\right)}_n(\cos(\omega x))=\sum_{k=0}^{\lfloor n/2\rfloor}(-1)^k\frac{\Gamma\left(\frac{\sqrt{5}}{2}+n-k\right)}{\Gamma\left(\frac{\sqrt{5}}{2}\right)k!(n-2k)!}(2\cos(\omega x))^{n-2k},
\ee 
and write
\be
c_n= \hat{c}_n \frac{\Gamma\left(\frac{\sqrt{5}}{2}\right)n!}{2^{n}\Gamma\left(\frac{\sqrt{5}}{2}+n\right) },\quad \hat{c}_n\ll 1.
\ee
Let us consider the following sum $\tilde{S}$
\be
\tilde{S}=\sum_{n=0}^{\infty}  \hat{c}_n \c_n, \quad \c_n=\left[\sin(\omega x)\right]^{\frac{1}{4}(\sqrt{5}-1)} \frac{\Gamma\left(\frac{\sqrt{5}}{2}\right)n!}{2^{n}\Gamma\left(\frac{\sqrt{5}}{2}+n\right) }C^{\left(\frac{\sqrt{5}}{2}\right)}_n(\cos(\omega x)), \quad |\c_n(x)|<1.
\ee
Convergence of the sum is ensure by choosing $\hat{c}_n=\hat{\c}_0(n+1)^{-s}$ with $\hat{\c}_0\ll1$, $s=\s+it$ and thus the series corresponds to a Dirichlet series, for details see for example 
  \cite{HardyR,helson1963,mandelbrojt2012dirichlet,apostol1998introduction}. Then, we obtain 
\be 
\tilde{S}(s) = \hat{\c}_0\sum_{n=1}^{\infty}\frac{\c_{n-1}}{n^s}.
\label{Dirichletsum}
\ee
For the convergence of the Dirichlet series we follow \cite{apostol1998introduction}. Let us consider $\s\geq a$, then $|n^s|=n^{\s}\geq n^a$ and
\be
\left|\frac{\c_{n-1}}{n^s}\right|\leq \frac{1}{n^a}. 
\ee
Now, if the Dirichlet series converges absolutely for $s=a+ib$ then it also converges absolutely for all $s$ with $\s\geq a$. For our case, the sum Eq. \eqref{Dirichletsum} converges absolutely for $\s> a= 1$ since the Riemann zeta function $\zeta(a)$ converges absolutely for $a> 1$. Hence, the fluctuations can be parametrized  as
\be
\vec{\zeta}_{\perp} =
\begin{pmatrix}
\sin(\omega x)\\
\cos(\omega x)
\end{pmatrix}
\left[\sin(\omega x)\right]^{\frac{1}{4}(\sqrt{5}-1)}\sum_{n=0}^{\infty} \frac{\hat{\c}_0}{(n+1)^a}\frac{\Gamma\left(\frac{\sqrt{5}}{2}\right)n!}{2^{n}\Gamma\left(\frac{\sqrt{5}}{2}+n\right) }C^{\left(\frac{\sqrt{5}}{2}\right)}_n(\cos(\omega x)).
\ee
with $a>1$. Finally, we can explicitly check that $\frac{\ud\vec{\phi}}{\ud x}\cdot\vec{\zeta}_{\perp}=0$. This is in agreement with the orthogonality condition stated at section 2. 

It remains to study the vertical line geodesics. The lines are described by $\phi^1(x)=\phi^1_0\ue^{mx}$, $\phi^2(x)=\phi^2_0$, where we set $m>0$. We use the general result of Example 1 and obtain the associated spectral problem for the vertical lines 
\be
 -\ddot{u}^{1}_{n}=\la_n u^1_{n}, \quad  -\ddot{u}^{2}_{n}+m^2u^2_{n}=\la_n u^2_{n}
\label{UeqVerLInes}   
\ee
For these geodesics, the fluctuations are 
\bea
\zeta^1_{\perp}(x)&=&\sum_{n}^{\infty}\left(a_n\cos(\sqrt{\la_n}x)+b_n\sin(\sqrt{\la_n}x)\right),\\
\zeta^2_{\perp}(x)&=&\sum_{n}^{\infty}\left( c_n\cos(\sqrt{\la_n-m^2}x)+d_n\sin(\sqrt{\la_n-m^2}x)\right).
\eea
The orthogonality condition $\frac{\ud\vec{\phi}}{\ud x}\cdot\vec{\zeta}_{\perp}=0$ implies that $a_n=b_n=0$. It remains to impose boundary conditions and to do so,  we introduce the regulator $R$. This is required since the endpoints of the straight line are at $x=\pm\infty$. The possible boundary conditions are:  Dirichlet boundary conditions $\zeta^2_{\perp}(-R)=0=\zeta^2_{\perp}(R)$ or mixed boundary conditions $\zeta^2_{\perp}(-R)=0=\left.\frac{\ud\zeta^2_{\perp}(x)}{\ud x}\right|_{x=R}$. The first possibility  is equivalent to put the problem in a ``box'' as is done for a free particle in quantum mechanics. The ``box'' has length $2R$ and  it is centred at $x=0$. Then the fluctuation can be written in term of the functions 
\be
 f_{n}^{(-)}(x)=\sin\left(\frac{\pi n}{R}x\right),\quad  f_{n}^{(+)}(x)=\cos\left(\frac{\pi (n-1/2)}{R}x\right),\quad n=1,2,3\ldots
\ee
with their respectively eigenvalues $\la_{(-)n}=(\pi n/R)^2+m^2$ and  $\la_{(+)n}=(\pi (n-1/2)/R)^2+m^2$. For  mixed boundary conditions,  the functions are 
\be
f_{n}^{(-)}(x)=\sin\left(\frac{\pi (n-1/2)}{2R}(x+R)\right),\quad  f_{n}^{(+)}(x)=\cos\left(\frac{\pi (n-1/2)}{2R}(x-R)\right),\quad
\ee
where $ n=1,2,3\ldots$ and their respectively eigenvalues $\la_{(-)n}=(\pi(n-1/2)/2R)^2+m^2=\la_{(+)n}$. We note that in the continuum limit $R\to\infty$,  we are require to take $n\to \infty$ in order to obtain other eigenvalues rather than $m^2$ in both cases. In summary, the fluctuations  are
\be
\zeta^2_{\perp}(x)=\sum_{n=1}^{\infty}\left[c^{-}_nf_{n}^{(-)}(x)+c^{+}_nf_{n}^{(+)}(x)\right].
\ee
Applying the knowledge  learned from the fluctuations for the semicircles,  we choose $c^{\pm}_n=\hat{\c}_0^{\pm}n^{-a}$ with $a>1$ and $\hat{\c}_0^{\pm}\ll 1$. 

We conclude with Table \ref{tablespectrumO1} of the spectra, without any zero modes, of the fluctuation operator  for the different geodesics.
\begin{table}[ht!]
\centering
 \begin{tabular}{l|l} 
\hline 
 Geodesic & Eigenvalue \\
 \hline
Semicircle& $\la_n=n(n-1+\sqrt{5})\omega^2+\frac{\omega^2}{4}(1-\sqrt{5})^2$, $\omega>0$, $n\geq 0$\\
Vertical line (Dirichlet $(+)$)& $\la_{(+)n}=(\pi (n-1/2)/R)^2+m^2$, $n\geq 1$\\
Vertical line (Dirichlet $(-)$)& $\la_{(-)n}=(\pi n/R)^2+m^2$, $n\geq 1$\\
Vertical line (Mixed $(+)$)& $\la_{(+)n}=(\pi(n-1/2)/2R)^2+m^2$, $n\geq 1$ \\
Vertical line (Mixed $(-)$)& $\la_{(-)n}=(\pi(n-1/2)/2R)^2+m^2$, $n\geq 1$\\
\hline
 \end{tabular}
 \caption{Spectra of the operator $\cO'$ for the geodesics of the two dimensional Poincar\'e half plane.}
 \label{tablespectrumO1}
\end{table}

\subsubsection{Zero modes}
We turn to the study of zero modes. First notice that if $\vec{u}_*$ is a zero mode then $P\vec{u}_*$ is also a zero mode. Therefore, we focus  on the zero modes in Eq. \eqref{UeqSemCir} and Eq. \eqref{UeqVerLInes}. For the vertical line with $G(x_0)=\pi/2$,  we find
\be
 \vec{u}_*=\begin{pmatrix}
 1\\
 0
 \end{pmatrix} \Rightarrow \vec{\chi}_0= \begin{pmatrix}
 0\\
 1
 \end{pmatrix}
\ee
and for $G(x_0)=-\pi/2$, the vertical line
\be
 \vec{u}_*=\begin{pmatrix}
 0\\
 \ue^{m x}
 \end{pmatrix} \Rightarrow \vec{\chi}_0= \begin{pmatrix}
 \ue^{m x}\\
 0
 \end{pmatrix}
\ee
As discussed in section 3, these zero modes are associated with the invariance under translations along the $\phi^2$-axis and dilatations of the background metric. Similarly, for the semicircle, we find that for $G(x_0)=-\omega x_0-\pi/2$ and	
\be
 \vec{u}_*=\begin{pmatrix}
 -1\\
 0
 \end{pmatrix} \Rightarrow \vec{\chi}_0= \begin{pmatrix}
 \sin(\omega x)\\
 \cos(\omega x)
 \end{pmatrix}.
\ee
The second zero mode is obtained by taking the limit $\omega\to 0$, i.e. 
\be
 \vec{\chi}_0= \begin{pmatrix}
 0\\
 1
 \end{pmatrix}.
\ee

\subsection{Quantum fluctuations}
\subsubsection{Relation with the Euclidean Wheeler-DeWitt equation}
The Euclidean Wheeler-DeWitt equation for the worldline is
\be
 \left[\frac{\pa^2}{\pa (\phi^1)^2}+\frac{\pa^2}{\pa (\phi^2)^2}+f^2(\phi^1,\phi^2)\frac{M^2}{\hbar^2}\right]\Psi(\phi^1,\phi^2)=0,
 \label{WdWeq}
\ee
where we have used Eq. \eqref{metricd2}. The corresponding equation for the Poincar\'e half plane is 
\be
 \left[\frac{\pa^2}{\pa (\phi^1)^2}+\frac{\pa^2}{\pa (\phi^2)^2}+\frac{1}{\hbar^2}\frac{L^2M^2}{(\phi^1)^2}\right]\Psi(\phi^1,\phi^2)=0,
 \label{WdWeq2}
\ee
We see that for 
\be
\frac{L^2M^2}{\hbar^2}\to -m_{\Phi}^2L^2, 
\label{Adscftscalarcontinuation}
\ee
the Eq. \eqref{WdWeq2} corresponds to the wave equation of a scalar field $\Phi$ with mass $m_{\Phi}$ in the two-dimensional hyperbolic background. The solution of the wave equation for the scalar field has been widely studied in the context of the AdS/CFT correspondence, see for example \cite{Witten:1998qj,Freedman:1998tz}. Here instead we are interested in the semiclassical approximation of the wavefunction. Therefore, as discussed in the introduction,   we write $\Psi=\exp\left(-\cS/\hbar\right)$. The equation for $\cS$ is
 \be
 \left(\frac{\pa\cS}{\pa\phi^1}\right)^2-\hbar\frac{\pa^2\cS}{\pa (\phi^1)^2}+\left(\frac{\pa\cS}{\pa\phi^2}\right)^2-\hbar\frac{\pa^2\cS}{\pa (\phi^2)^2}+\frac{L^2M^2}{(\phi^1)^2}=0.
 \label{SEsc}
\ee
The solution at all orders in $\hbar$ of Eq. \eqref{SEsc} is
\bea 
\cS_q=\cS_0+q(\a_{\pm}\ln(\phi^1)+\b\ln(\phi^2)), \quad \a_{\pm}=-\frac{q\hbar }{2}\pm iLM\sqrt{1-\frac{\hbar^2}{4L^2M^2}},
 \eea
where $q=\pm 1$, $ \b(\b+q\hbar)=0$ and $\cS_0$ is a constant. We choose $\b=0$ and  obtain
\bea 
\cS(\phi^1)=\cS_0+\left(-\frac{\hbar }{2}\pm iLM\sqrt{1-\frac{\hbar^2}{4L^2M^2}}\right)\ln\phi^1
\label{solutionofS}
 \eea
The wavefunction for $\b=0$ in the semiclassical limit is
\be
\lim_{\hbar\to 0}\Psi =\cA\sqrt{\phi^1}\ue^{\pm\frac{iLM}{\hbar}\ln\phi^1},
\label{Semiclaswavefuncworldline} 
\ee
where $\cA$ in the normalization constant. Alternatively, if we consider 
\be 
\Psi=\ue^{-i\frac{\omega}{\hbar}\phi^2}\psi(\phi^1), \quad \omega\geq 0,
\ee
then Eq. \eqref{WdWeq2} becomes the Schr\"odinger equation for  bound states 
\be
-\hbar^2\frac{\ud^2\psi}{\ud(\phi^1)^2}+V\psi=E\psi, \quad V=-\frac{M^2L^2}{(\phi^1)^2},\quad E=-\omega^2.
\label{eqforpsi}
\ee
The WKB approximation of this equation for $E=0$ gives Eq. \eqref{Semiclaswavefuncworldline}. The approximation holds if we are far way from the classical turning point which is located at $\phi^1\to \infty$. Therefore Eq. \eqref{Semiclaswavefuncworldline} holds for $\phi^1\ll\infty$.

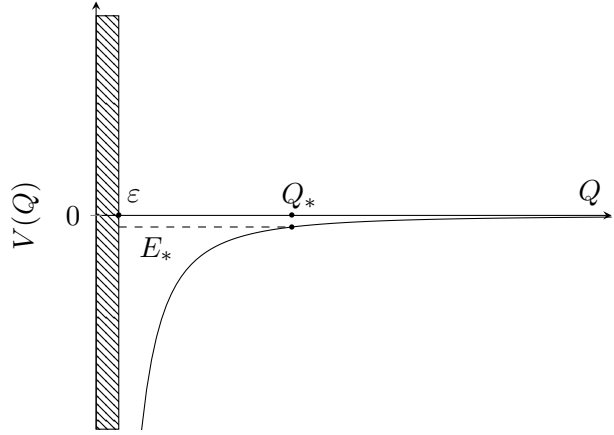
\begin{figure}[ht!]
\centering
\begin{tikzpicture}
\draw[pattern=north west lines] (0,0) rectangle (0.3,5.5);

 \begin{axis}[domain  = 0:4,
			    axis lines = middle,
               samples = 100,
               xmin    = 0,
               xmax    = 4,
               ymin    = -8,
               ymax    = 8,
               ytick   = \empty,
               xtick   = \empty,
               xlabel  = {$Q$},
               ylabel  = {$V(Q)$},
               extra y ticks = {0},
               ylabel near ticks,
              ]

\addplot [
    domain=0.2:4, 
    samples=100, 
]
{-1/(x*x)};
\end{axis}
\fill (2.6,2.85)  circle[radius=1pt];
\node at (2.7,3.1) {$Q_{*}$};
\fill (2.6,2.69)  circle[radius=1pt];
\draw[dashed] (0.3,2.69)--(2.6,2.69);
\node at (0.8,2.4) {$E_{*}$};
\fill (0.3,2.85)  circle[radius=1pt];
\node at (0.5,3.1) {$\varepsilon$};

%
%
%
%
\end{tikzpicture}
\caption{Regularization of the conformal quantum mechanics model.}
\label{RegularizationofCQM}
\end{figure}

We have yet to discuss divergences at $\phi^1\to 0$ and $\phi^1\to \infty$ for general solutions of Eq. \eqref{solutionofS}. This issue can be tackled by noticing that Eq. \eqref{eqforpsi} can be derived from  a quantum mechanical model,  whose action is  
\be
I= \frac{1}{2}\int\ud t\left(\dot{Q}^2-V\right),\quad V=-\frac{g}{Q^2},
\label{CQMaction}
\ee
where $g=M^2L^2$, $t\leftrightarrow \phi^2$ and $Q\leftrightarrow\phi^1$. The system described by this action equals to the  Euclidean  conformal quantum mechanics model developed in \cite{deAlfaro:1976vlx}. The action is invariant, up to a boundary term, under
\be
t\to \frac{at+b}{ct+d},\quad Q(t)\to\frac{Q(t)}{ct+d}, 
\ee
with $a,b,c,d\in\mathbb{R}$ and $ad-bc=1$ and the  one-dimensional conformal group $SL(2,\mathbb{R})$ is a symmetry of the action. To ensure finiteness of the solution, we introduce a length scale $\varepsilon>0$ and a energy scale $E_{*}<0$. Therefore, we explicitly break conformal invariance. The regulated Hamiltonian is defined to be
\be
H=\frac{1}{2}\left(P^2+V\right),\quad V=\left\{\begin{matrix}
\infty & 0<Q<\varepsilon\\
-\frac{g}{Q^2}& Q>\varepsilon
\end{matrix}
\right. .
\ee
In Figure \ref{RegularizationofCQM} we show the vertical wall at $Q=\varepsilon$ and the energy $E_{*}=-\frac{g}{2Q_*^2}$ where $Q_{*}$ is the location of the turning point. 
As discussed in \cite{griffiths2017introduction}, the WKB solution for a potential with a wall is of the form
\be
\psi(Q)=\frac{\cA}{\sqrt{P(Q)}}\sin\left(\frac{1}{\hbar}\varphi(Q,Q_*)+\frac{\pi}{4}\right),\quad  \varphi(Q,Q_*)=\int\limits_{Q}^{Q_*}\ud q \,P(q)  ,\quad Q\ll Q_*, 
\ee
where  $P(Q)=\sqrt{2E-V(Q)}$ is the classical momentum and $\varphi(\varepsilon,Q_*)=\left(k-\frac{1}{4}\right)\hbar \pi$  for $k=1,2,\ldots$. The later condition ensures that $\psi(\varepsilon)=0$.
Hence, the regulated wavefunction is
\be
 \psi(Q)=\cA\sqrt{Q}\sin\left(\frac{1}{\hbar}\varphi(Q,Q_*)+\frac{\pi}{4}\right),\quad  Q\ll Q_*, 
 \label{Semiclaswavefuncworldline2}
\ee
with 
\be
 \varphi(Q,Q_*)=\sqrt{g}\left[\ln\left(\frac{Q}{Q_*}\right)+\ln\left(\sqrt{1-\left(\frac{Q}{Q_*}\right)^2}+1\right)-\sqrt{1-\left(\frac{Q}{Q_*}\right)^2}\right].
%
\ee
This result is in agreement with Eq. \eqref{Semiclaswavefuncworldline} by considering the Taylor expansion of $\varphi(Q,Q_*)$ for small $Q/Q_*$ in Eq. \eqref{Semiclaswavefuncworldline2}.

\subsubsection{Regularization methods}
Now that the semiclassical solution of the Wheeler-DeWitt equation is fully understood, we proceed to compare the one-loop correction of this solution with the equivalent correction via the path integral
\be
\sqrt{Q}\Leftrightarrow \frac{1}{\sqrt{\det\cO}}, 
\ee
where the spectrum of $\cO$ was developed in section \ref{PHPsection}. Recall that the Fadeev-Popov for the normal gauge is absorbed by the normalization constant of the path integral. The relation of the divergences with both equivalent expressions is 
\bea 
\prod_n^{\infty}\la_n\to 0\quad\mathrm{or}\quad \sum_n^{\infty}\ln\la_n\to -\infty & \Longleftrightarrow & Q\to\infty,\\
\prod_n^{\infty}\la_n\to \infty\quad\mathrm{or}\quad\sum_n^{\infty}\ln\la_n\to \infty &\Longleftrightarrow & Q\to 0.
\label{Q0divergence}
\eea
Our goal is to study regularization methods for the divergence given in Eq. \eqref{Q0divergence}. First, the cut-off regularization employed in the Wheeler-DeWitt approach implies
\be
\sum_n^{N}\ln\la_n \quad \Longleftrightarrow \quad Q\to \epsilon,
\ee
where $N$ is large but finite. This is the most straightforward method to regularize the sum.
To study all the cases at once, we  parametrize the eigenvalues as
\be
\la_n=c^2\left[(n-a)^2+b^2\right], \quad n=1,2,3,\ldots,
\label{eigenparamallcasesR}
\ee
and the cases are listed in Table \ref{tablespectrumO1}. For the semicircle geodesics, we have $a=(3-\sqrt{5})/2\approx 0.3820$, $c=\omega>0$, $b=0$ and for the straight geodesics,  $a=0,1/2$, $c=\pi/(2R),\pi/R$ and $b=m/c$. The spectral function of interest is the zeta function associated to the operator $\cO$,
\be
 \zeta(s,\cO)=\sum_{n=1}^{\infty}\frac{1}{(L^2\la_n)^s},
\label{zetafunctiondef}
\ee
where  $\la_n$ is defined in Eq. \eqref{eigenparamallcasesR}. The sum is absolutely convergent for $\Re{s}>B$ for a constant $B>0$. The radius of the half plane $L$ is introduced in order to make the expression dimensionless. The relation between the zeta function  with the sum in question is given by the formal expression
\be
-\zeta'(0,\cO)=\sum_{n=1}^{\infty}\ln(L^2\la_n).
\ee 
For the eigenvalues of interest,  we have $0\leq a <1/2$ and to focus on the large $n$ behaviour of the sum, we set $a=0$ in Eq. \eqref{zetafunctiondef}. For large $n$, we have
\be
-\zeta'(0,\cO)\approx 2Lc\sum_n 1+2\sum_n\ln n\to \infty.
\ee
We can regularize this expression using a hard cutoff $\sum_{n}^N$ and thus for large $N$, we get $-\zeta'(0,\cO)\sim -2\ln(R/ L)N +2\ln N $. Using Eq. \eqref{eigenparamallcasesR} we write the approximation of the zeta function as
\be
 \zeta(s,\cO)\approx\frac{1}{2(cL)^{2s}}\left[\sum_{n=-\infty}^{\infty}\frac{1}{(n^2+b^2)^s}-\frac{1}{b^{2s}}\right], \quad b\neq 0.
\label{zetafunctiondefapprox}
\ee
By means of the Poisson summation formula the sum gives
\be
 \sum_{n=-\infty}^{\infty}\frac{1}{(n^2+b^2)^s}=\frac{\sqrt{8\pi b}}{\Gamma(s)}\ue^{-s\ln(2b)}\left[2\sum_{n=1}^{\infty}(2\pi n)^{s-1/2} K_{s-1/2}(2\pi n b)+\tilde{\epsilon}^{s-1/2}K_{s-1/2}(\tilde{\epsilon} b)\right],
\ee
where $K_{\nu}$ is the modified Bessel function of the second kind and $\tilde{\epsilon}$ is a cut-off since $K_{\nu}(0)$ diverges. Dropping this artificial divergence, we obtain
 \be
-\zeta'(0,\cO)\approx -\ln(bcL)+\ln\left(1-\ue^{-2\pi b}\right).
\label{sumregulatedzeta}
\ee
Recall that for the vertical line geodesic the parameters are $bc=m$ and $b=(mR)/\pi,(2mR)/\pi$. Therefore, as $R\to \infty$ then $\sum_{n=1}^{\infty}\ln(L^2\la_n)\approx -\ln(mL)$. 

For the semicircle geodesic, we consider $ \zeta(s,\cO)\approx\sum_{n=1}^{\infty}1/(\omega L n)^{2s}$. Hard cut-off regularization gives $-\zeta'(0,\cO)\sim 2\ln(\omega L)N +2\ln N $. In order to connect with the approximation of the zeta function for the vertical line geodesic by means of the Poisson summation formula, we consider the following regularization
\be
\frac{1}{n^2}\to\frac{1}{n^2+N^2}.
\ee
Thus, we obtain Eq. \eqref{sumregulatedzeta} with $b=N\omega$ and $c=\omega$. This implies that for large $N$ the sum goes as $-\zeta'(0,\cO)\sim -\ln N -\ln(\omega L)$. 
 
Another route to compute $\zeta'(0,\cO)$ uses the relation of the zeta function with the trace of the heat kernel $K(\tau)=\sum_{n=1}^{\infty}\exp (-\la_n \tau)$, see \cite{Vassilevich:2003xt,Fursaev:2011zz}. The relation is given by the Mellin transform
\be
\zeta(s,\cO)=\frac{1}{L^{2s}\Gamma(s)}\int\limits_{0}^{\infty}\ud \tau\, \tau^{s-1}K(\tau),
\ee
where $\tau$ has dimensions of (mass)$^{-2}$. The asymptotic expansion of $K(\tau)$ of our problem can be computed using recent results developed in \cite{10.2307/41291944,MaoRen2013}.  Using the expansion of the auxiliriay function $G_2$ given in Eq. (1.4) reported in \cite{MaoRen2013}, the trace of the heat kernel for $c^2 \tau\to 0$ and $a=0$ is\footnote{This result is based on the Euler-Maclaurin summation formula for the  asymptotics of the sum developed in \cite{Dzagier006}.}
\be
K(\tau)\approx-\frac{1}{2}\ue^{-c^2b^2\tau}+\frac{\sqrt{\pi}}{2c}\frac{\ue^{-c^2b^2 \tau}}{\sqrt{\tau}}.
\label{RamanujanKexp}
\ee
The same result can be obtained by applying the Poisson summation formula as shown in section 4.2 in \cite{Fursaev:2011zz}. These two methods of computing the high energy expansion ($\tau\to 0$) of $K(\tau)$ do not involve any information of the boundary conditions from which the spectrum $\{\la_n\}$ is obtained. The boundary conditions are taken into account in the untraced heat kernel $K(x,x';\tau)$ defined on $\Sigma\times\Sigma\times\mathbb{R}^+$. The simplest way to treat boundary conditions in one dimension is to use the fact that an even function,  $f_+(x)=(f(x)+f(-x))/2$, satisfies Neumann-type boundary conditions at $x=0$ and an odd function  $f_-(x)=(f(x)-f(-x))/2$ satisfies Dirichlet-type boundary conditions  at $x=0$. This implies that the untraced heat kernel in the presence of boundaries in $\Sigma$ can be written as
\be
 K_{\pm}(x,x';\tau)=\frac{1}{2}(K(x,x';\tau)\pm K(-x,x';\tau)),
\ee
where $K_{\pm}$ corresponds to the symmetrization $(+)$ or antisymmetrization $(-)$ of the heat kernel. Therefore, the eigenfunction expansion results in
\be
 K_{\pm}(x,x';\tau)=\frac{1}{2}\sum_n\ue^{-\la_n \tau}\d_{AB} (\chi_n^A(x)\pm \chi_n^A(-x))\chi^B_n(x').
\ee
It is clear that if the eigenfunctions are odd we are dealing with Dirichlet-type boundary conditions and for even eigenfunctions we have Neumann-type boundary conditions. For our problem, the fluctuations for the semicircle geodesic are set to vanish at the endpoints ($\pa\Sigma=\{0,\pi/\omega\}$) and similarly for the vertical geodesic ($\pa\Sigma=\{-R\}$ or $\pa\Sigma=\{-R,R\}$). This implies that we have used  Dirichlet-type boundary conditions and we expect that the untraced heat kernel must be antisymmetric. Indeed, this is the case since Eq. \eqref{RamanujanKexp} can be derived from the antisymmetrization of the free heat kernel of the operator $-\ud^2/\ud x^2+ b^2c^2$ defined on the interval $[0,\Delta L]$. After taking the trace  we obtain for $\tau \to 0$
\be
2K_-(\tau)=\frac{\ue^{-b^2c^2\tau}}{\sqrt{4\pi\tau}}\Delta L-\frac{\ue^{-b^2c^2\tau}}{2}.
\ee
For $c=\pi/\Delta L$ we recover Eq. \eqref{RamanujanKexp}, this result is discussed in section 4.5 in \cite{Fursaev:2011zz}.

Returning to the computation of $-\zeta'(0,\cO)$,  we integrate the trace of the heat kernel from  $\tau_{UV}$ to $\tau_{IR}$ and find
\be
-\zeta'(0,\cO)\approx 
\left\{
\begin{array}{ll}
\left[\frac{1}{2}\mathrm{Ei}(-b^2c^2 \tau)+\pi b\,\mathrm{erf}(bc\sqrt{\tau})+\frac{1}{c}\sqrt{\frac{\pi}{\tau}}\ue^{-b^2c^2\tau}\right]_{\tau_{UV}}^{\tau_{IR}} & b\neq 0,\\
\left[\frac{1}{c}\sqrt{\frac{\pi}{\tau}}+\frac{1}{2}\ln \tau\right]_{\tau_{UV}}^{\tau_{IR}} &b=0,
\end{array}
\right.
\ee
where Ei corresponds to the exponential integral. Let us introduce the dimensionless parameters
\be
\tilde{\tau}_{IR}=\frac{\tau_{IR}}{L^2}, \quad \tilde{\tau}_{UV}=\frac{\tau_{UV}}{L^2}. 
\ee
Since $\tilde{\tau}_{IR}\sim 1$ (recall that the expansion of $K(\tau)$ is for small $\tau$) and $\tilde{\tau}_{UV}\ll 1$, the divergences are
\be
-\zeta'(0,\cO)\approx 
\left\{
\begin{array}{ll}
-\frac{1}{cL}\sqrt{\frac{\pi}{\tilde{\tau}_{UV}}}-\frac{1}{2}\mathrm{Ei}(-b^2c^2L^2 \tilde{\tau}_{UV})& b\neq 0,\\
-\frac{1}{cL}\sqrt{\frac{\pi}{\tilde{\tau}_{UV}}}+\frac{1}{2}\ln \tilde{\tau}_{UV} &b=0.
\end{array}
\right.
\ee

Instead of computing the zeta function, we can directly compute Eq. \eqref{sumregulatedzeta}. For the vertical line geodesic and using the Euler-Maclaurin formula\footnote{The Euler-Maclaurin formula is 
\be 
\sum _{n=a}^{N}\la_n\approx \int\limits _{a}^{N}\ud n\, \la(n)+{\frac {f(N)+f(a)}{2}}+\sum _{k=1}^{N }\,{\frac {B_{2k}}{(2k)!}}\left(f^{(2k-1)}(N)-f^{(2k-1)}(a)\right),
\ee
where $B_k$ are the Bernoulli numbers.
} 
for large  $N$ and fixed $R$, we obtain
\be
\sum_{n=1}^{\infty}\ln(L^2\la_n)\approx 2N\ln N -2N\ln(R/L) +qmR,\quad q=1,2,
\ee
and for the semicircle geodesic
\be
\sum_{n=1}^{\infty}\ln(L^2\la_n)\approx 2N\ln N +2N\ln(\omega L).
\ee
\begin{table}[!htb]
      \centering
        \begin{tabular}{l l | l}
        \hline 
            &Regulator & Divergences for the vertical line geodesic\\
            \hline
            1)&$I$ and hard-cutoff& $I\sim -2N\ln(R/L) +2\ln N$\\
            2)&$I=-\zeta'(0,\cO)$ and Poisson & $I\sim\ln\left(1-\ue^{-4mR}\right)$\\
            3)&$I=-\zeta'(0,\cO)$ and $K(\tau)$& $I\sim-\frac{qR}{\pi L}\sqrt{\frac{\pi}{\tilde{\tau}_{UV}}}-\frac{1}{2}\mathrm{Ei}(-m^2L^2 \tilde{\tau}_{UV})$, $q=1,2$\\
            4)&$I$ and Euler-Maclaurin & $I \sim 2N\ln N -2N\ln(R/L) +qmR$, $q=1,2$\\
            \hline 
            \hline
            &Regulator &Divergences for the semicircle geodesic\\
            \hline
            5)&$I$ and hard-cutoff& $I\sim 2N\ln(\omega L) +2\ln N$\\
            6)&$I=-\zeta'(0,\cO)$, $\frac{1}{n^2}\to\frac{1}{n^2+N^2}$ and Poisson & $I\sim -\ln N$\\
            7)&$I=-\zeta'(0,\cO)$ and $K(\tau)$& $I\sim -\frac{1}{\omega L}\sqrt{\frac{\pi}{\tilde{\tau}_{UV}}}+\frac{1}{2}\ln \tilde{\tau}_{UV}$\\
            8)&$I$ and Euler-Maclaurin & $I\sim 2N\ln N +2N\ln(\omega L)$\\
            \hline
        \end{tabular}
  \caption{Regulators and divergences for $I\equiv\sum_{n=1}^{\infty}\ln(L^2\la_n)$.}
\label{divtable}
\end{table}
The regulators and divergences are listed in Table \ref{divtable}. We see that the for $\tilde{\tau}_{UV}\to 0$ or $N\to\infty$ the cases 3), 6) and 7) break the semiclassical approximation. To relate the computation of $I$ with the Euler-Maclaurin formula, the zeta function and $K(\tau)$, let us define the integral representation
\be
K(\tau)\equiv \int\limits_{v_-}^{v_+}\ud v\, \rho(v)\ue^{-v^2 \tau},
\label{intKrepv}
\ee
where $\rho(v)$ corresponds to the density of states\footnote{The proposed integral representation of $K(\tau)$ follows the same line of \cite{Dai:2009zza}. In this reference, the function $\rho$ is referred as to the density of states and  is defined as
\be
N(\la)=\sum_{\la_n<\la}1 =\int\limits_0^{\la}\ud \la'\, \rho(\la'),
\ee
where $N(\la)$ is the counting function. This function is defined as the number of eigenvalues smaller than $\la$. On the other hand, we can interpret the integral representation as the high temperature limit of a partition function. Let 
\be
\epsilon_n =L\la_n=L((cn)^2+(cb)^2),\quad \b = \frac{\tau}{L}=L\tilde{\tau}=\frac{1}{k_B T}. 
\ee
Then $K(\beta)$ is a partition function of a system with energy eigenvalues $\epsilon_n$. In the high temperature limit $\beta \to 0$, we obtain
\be
 K(\beta)\approx \ue^{-L(bc)^2\beta} \int\limits_0^{\infty}\ud v \,\rho(v)\ue^{-Lv^2\beta},\quad v=nc,
\ee
which is the same form as the integral representation.\label{statmechsug}
}. Taking the integral in $\tau$ from $\tau_{UV}$ to $\tau_{IR}$, we obtain
\be
I\approx - \int\limits_{v_-}^{v_+}\ud v\, \rho(v)(\mathrm{Ei}(-v^2\tau_{IR})-\mathrm{Ei}(-v^2\tau_{UV})),
\label{Iapproxlargev}
\ee
where $\mathrm{Ei}$ corresponds to the exponential integral. For the determination of the function $\rho$, we consider the change of variable $v^2=u+u_0$ where $u_0$ is a constant. Then
\be
K(\tau)= \frac{\ue^{-u_0 \tau}}{2}\int\limits_{v^2_--u_0}^{v^2_+-u_0}\ud u\, \frac{\rho(u)}{\sqrt{u+u_0}}\ue^{-u \tau}.
\label{heatkernelenergyrep}
\ee
Let us assume that $\rho(u)\sim\sqrt{u+u_0}(u+u_0)^{\a-1}$ and $u\gg u_0$ (large $v$), thus we obtain
\be 
K(\tau)\sim\frac{\ue^{-u_0 \tau}}{2}\frac{\Gamma(\a)}{\tau^{\a}}, \quad \a>0,
\ee
for $v_-=\sqrt{u_0}$ and $v_+\to \infty$. For the case $\a=0$ we take $\rho(u)\sim\sqrt{u+u_0}\,\d(u)$, where $\d$ corresponds to the Dirac delta function, with the previous bounds. The asymptotic expansion given in Eq. \eqref{RamanujanKexp} can be recovered if the density of states is of the form
\be
 \rho(u)\approx\frac{1}{c}-\sqrt{u}\,\d(u), \quad\mathrm{or}\quad  \rho(v)\approx \frac{1}{c}-v\,\d (v^2).
 \label{roapprox}
\ee
The correspondence of  terms from the density of states to Eq. \eqref{RamanujanKexp}  are
\be
\frac{1}{c}\to \frac{\sqrt{\pi}}{2c}\frac{\ue^{-c^2b^2 \tau}}{\sqrt{\tau}}, \quad -\sqrt{u}\,\d(u)\to -\frac{1}{2}\ue^{-c^2b^2\tau}.
\ee
We can naively conclude that $ \rho\approx \frac{1}{c}$ gives the one-dimensional free heat kernel. As already discussed, Eq. \eqref{RamanujanKexp} corresponds to the antisymmetrized one-dimensional free heat kernel and therefore Eq. \eqref{roapprox} actually takes into account the antisymmetrization or equivalently it takes into account the presence of boundary conditions\footnote{This result can be generalized straightforwardly. Let $\cO_{1+p}+\mu^2$ be an operator defined in a $(1+p)$-dimensional manifold $\Sigma$ and $\mu$ a mass term. The high energy expansion ($\tau\to 0$) of the trace of the heat kernel can be written as
 \be
 K(\tau)\approx\ue^{-\mu^2\tau}\sum_{l=0}^{\infty} a_l\frac{1}{\tau^{(D-l)/2}}, 
 \ee
 where $a_{l}$ are the integrated heat kernel coefficients, see \cite{Vassilevich:2003xt,Fursaev:2011zz}. As stated in \cite{BransonGilkey90,Vassilevich:2003xt}, the half powers in $\tau$ arise from the presence of boundaries in $\Sigma$. In the above sum, the term for which $l=D$ is independently of $\tau$  and the divergent terms are given for $l<D$. This expression can be obtained from \eqref{heatkernelenergyrep} (taking $u\gg u_0=\mu$ and $v_+\to \infty$) via the following density function 
 \be
 \rho(u)\approx \sqrt{u}\sum_{l\neq D}^{\infty} 2a_l\frac{1}{\Gamma((D-l)/2)}u^{(D-l)/2-1}+2a_D \sqrt{u}\d(u).
 \label{densityrho(u)}
 \ee 
Taking $v^2\approx u$ and substituting the above result in \eqref{Iapproxlargev}, we obtain the terms expected from the  Euler-Maclaurin summation formula.}.

Continuing with the goal of relating the regularization methods, let us consider that the density of states is constant, i.e. $ \rho\approx \frac{1}{c}$. The integral given in Eq. \eqref{Iapproxlargev} for fixed $\tau_{UV}$ and $\tau_{IR}$  results in
\be
I\approx -\frac{1}{c}\left[v\left(-\mathrm{E}_1(v^2\tau_{IR})+\mathrm{E}_1(v^2\tau_{UV})\right)-\sqrt{\frac{\pi}{\tau_{IR}}}\mathrm{erf}(\sqrt{\tau_{IR}}v)+\sqrt{\frac{\pi}{\tau_{UV}}}\mathrm{erf}(\sqrt{\tau_{UV}}v)\right]^{v_+=\Lambda}_{v_-=bc},
\ee
where $\mathrm{E}_1(x)=-\mathrm{Ei}(-x)$ and $\Lambda\to \infty$. In terms of the dimensionless parameters $\tilde{\tau}_{IR}$, $\tilde{\tau}_{UV}$ and 
\be
N=\frac{\Lambda}{c},
\ee
we evaluate the result  for non-zero $b$  and $\sqrt{\tau_{UV}}N\gg 1$. The sum gives
\be
I\approx -N\mathrm{E}_1(N^2c^2L^2\tilde{\tau}_{UV})
-\frac{1}{cL}\sqrt{\frac{\pi}{\tilde{\tau}_{UV}}}
+b(-\mathrm{E}_1(b^2c^2L^2)+\mathrm{E}_1(b^2c^2L^2\tilde{\tau}_{UV}))+2b+\frac{\sqrt{\pi}}{cL}\mathrm{erfc}(bcL).
\ee
For the case $b=0$, we only consider the first two terms. Using the expansion
\be
 \mathrm{E}_1(x)=-\gamma-\ln x -\sum_{k}^{\infty}\frac{(-1)^k x^k}{k!k},
\ee
where $\c$ is the Euler-Mascheroni constant. The divergences for the vertical line geodesics are
\be
 I\approx N(\c+\ln\tilde{\tau}_{UV})-\frac{qR}{\pi L}\sqrt{\frac{\pi}{\tilde{\tau}_{UV}}}+\frac{qmR}{\pi}\mathrm{E}_1(m^2L^2\tilde{\tau}_{UV})+2N\ln N-2N\ln(R/L)+O(N^3),
\ee
where the fourth and fifth terms corresponds to the  divergences calculated from the Euler-Maclaurin summation formula. The divergences for the semicircle geodesics are
\be
 I\approx N(\c+\ln\tilde{\tau}_{UV})-\frac{1}{\omega L}\sqrt{\frac{\pi}{\tilde{\tau}_{UV}}}+2N\ln N+2N\ln(\omega L)+O(N^3),
\ee
where the third and fourth terms corresponds to the  divergences calculated from the Euler-Maclaurin summation formula.

Last but not least, we can further exploit the integral representation  given in Eq. \eqref{intKrepv} by relating $K(\tau)$ with the phase-shift method as discussed in \cite{Pang2012}. The density of states is generically defined as
\be
\rho(v)=\rho_0+\frac{1}{\pi}\frac{\ud \d}{\ud v}, 
\label{phaseshiftdensitystates}
\ee
where $\rho_0$ is a constant and $\d(v)$ is the phase-shift, see \cite{Schwinger1954}. The existence of the phase-shift is due to the scattering process in the presence of a potential term (with possible derivative terms) in the operator. The comparison of Eq. \eqref{roapprox} with Eq. \eqref{phaseshiftdensitystates}, suggest that $\rho_0=1/c$ and we have a phase shift $\d = -\pi/2$\footnote{We have 
\be 
\d = -\pi \int \ud v\, v\d(v^2) = -\frac{\pi}{2}.
\ee
}. This implies a tension since we have previously discussed that we are dealing with an antisymmetrized  free heat kernel, i.e. the $-v\d(v^2)$ contribution in Eq. \eqref{roapprox} is not due to a potential (it arises due to boundaries).  

This can be settled on physical terms by analyzing the solution of the Wheeler-DeWitt equation given in Eq. \eqref{Semiclaswavefuncworldline2}. We first write 
\be
\sin\left(\frac{1}{\hbar}\varphi(Q,Q_*)+\frac{\pi}{4}\right)=\sin\left(PQ+\d\right), 
\ee
thus the phase shift is given by
\be
\d =  \frac{1}{\hbar}\varphi(Q,Q_*)+\frac{\pi}{4}-\frac{PQ}{\hbar}.
\label{WKBphaseshift1d}
\ee
This phase is referred as to the WKB phase shift. Secondly, we follow \cite{Graham:2001iv} and consider the summation rule
\be
\int\limits_0^{\infty}\ud k \frac{k^{2n}}{\pi}\frac{\ud}{\ud k}\left(\d (k)-\sum_{\nu=1}^m\d^{(\nu)}(k)\right) =\sum_j (-\kappa^2_j)^{n},
\ee
where $j$ runs over bound sates and binding energy $-\kappa^2_j$ and $\d^{(\nu)}$ is the $\nu$-th Born approximation. For $n=m=0$, we recover  Levinson's theorem. In order to obtain the WKB approximation of the summation rule, we expand Eq. \eqref{WKBphaseshift1d} for small $E_*$ 
\be
 \d =  \frac{1}{\hbar}\int\limits_Q^{Q_*}\ud q\sqrt{-V(q)}(1-\frac{E_*}{V(q)}+O(E_*^2))+\frac{\pi}{4}-\frac{PQ}{\hbar},
\ee
where the $\nu$-th Born approximation to the phase shift correspond to the term of order $(-V)^{\nu}$. After this identification, it is stated in \cite{Graham:2001iv} that the WKB approximation of Levinson's theorem is of the form
\be
\sum_j 1\approx \frac{2}{\pi} \int\limits_Q^{Q_*}\ud q\sqrt{-V(q)}=\frac{2}{\pi}\sqrt{\frac{g}{2}}\ln\left(\frac{Q_*}{Q}\right).
\ee
For our potential, the limit $Q\to 0$ implies that there are infinite number of bound states. This conclusion is in agreement with \cite{doi:10.1119/1.2165248}. The wall regularization $\epsilon \leq Q\ll Q_*$ allows, beside the already given bound states, the existence of a ground state and  well-defined phase shifts for the scattering states. Moreover, \cite{doi:10.1119/1.2165248} shows that for $\epsilon\to 0$ the phase shift oscillates between $\pm \pi/4$. If we further assume that $ML\to1/2$ the phase shift is $\pi/4$ and there is only one bound state. 

Therefore, the phase shift naively computed from the heat kernel expansion does not enter in the previous physical analysis. More precisely, the coefficient in the high energy expansion of the heat kernel that corresponds to a term of the form $v\d(v^2)$ in the density of states, does not contribute to the phase shift. Thus we should expect that sub-leading terms in the expansion of the heat kernel must take into account the WKB expression of the phase shift. The mathematical resolution of our naive computation requires the use the algebraic property of the Dirac delta function: $ \d(v^2-a^2)=\frac{1}{2|a|}(\d(v-a)+\d(v+a))$. Then
\be
\d =-\frac{\pi}{2}\lim_{a\to 0}\frac{1}{|a|}\int\ud v\, v(\d(v-a)+\d(v+a)) =0.
\label{diracandphase}
\ee 

We  conclude that computing a functional determinant via the heat kernel method is very useful not only due to its covariant nature but because it allow us to connect with other methods. Together with the information provided from the solution of the Wheeler-DeWitt equation, we can give a more precise answer to the one-loop correction to the wave-function of the system. Moreover, this can be discussed in a general framework by considering a statistical mechanics auxiliary system\footnote{In fact this was already suggested in the footnote    
 \ref{statmechsug} in page 28.}. Let us define the quantum Hamiltonian for this system as $\mathscr{H}=L\cO$ and the inverse temperature as $\b=\tau/L$. Then the trace of the heat kernel corresponds to the partition function of this auxiliary system, i.e. $K(\b L)\equiv \cZ(\b)=\Tr(\ue^{-\mathscr{H}\b})$. This is in essence the same as the procedure developed in \cite{Mukhanov:2007zz} and the path integral representation of this partition function led us to the worldline formalism, see for example \cite{Bastianelli:2005rc}. In this formalism, we construct a classical Hamiltonian from its quantum counterpart $\mathscr{H}=L\cO$ in the Schr\"odinger picture by identifying $\pi_{\mu}\leftrightarrow -i\pa_{\mu}$. 
 
From Eq.\eqref{THEOPERATOROFFLUC}, we see that the general form of the classical Hamiltonian follows from the background geometry. For a flat  background, the spin connection vanishes and the classical Hamiltonain contains a kinetic term proportional to $\pi_{\mu}^2$ and a potential term from the extrinsic curvature which depends on the background fields $\bar\phi^i$. For the hyperbolic background the kinetic term is also present but the spin connection and curvature induces new interactions. An explicit auxiliary statistical system is developed in the next section.
 
In terms of the untraced heat kernel, we can write $\cZ(\b)=\mu\int_{\Sigma}\ud x\, K(x,x;\b L)$ where  $K(x,x';\b L)=\left\langle x|\ue^{-\mathscr{H}\b}|x'\right\rangle$.  Then the high energy expansion of $K(x,x';\b L)$ corresponds to the high temperature expansion of $\cZ(\b)$. We expect that the high temperature expansion corresponds to a semiclassical description of the system. This can be argued by means of the thermal de Broglie wavelength defined as $\la_{\mathrm{th}}\sim \hbar \sqrt{L\b} $. The semiclassical description is obtained by requiring that $\la_{\mathrm{th}}$ must be smaller than any other characteristic length in the system. We fix $\hbar=1$ and  assume that the characteristic length is $L$. On physical grounds,  this is supported since in the tensionless limit the system behaves essentially like a free gas of massless particles and therefore the only characteristic lengths remaining arise from the background geometry. The semiclassical limit implies $\sqrt{\b/L}\ll 1$ and the high temperature limit satisfies this condition for $L$ fixed. Notice that if we consider the flat limit $L\to\infty$, one can find high temperatures such that the semiclassical condition is satisfied.

\section{Circular Wilson loop in Euclidean AdS$_3$}

The study of holographic Wilson loops dual to minimal surfaces in Euclidean AdS$_3$ has been carried out in \cite{Ishizeki:2011bf,Kruczenski:2013bsa,Kruczenski:2014bla,Irrgang:2015txa,Huang:2016atz,He:2017zsk,He:2017cwd,Cooke:2018obg}. An integrability-based and manifestly conformally invariant formalism  has provided analytic solutions that allow us to depart from the most well-studied holographic loop, i.e. the circular loop dual to a semi-sphere. The study of minimal surfaces in 3-dimensional hyperbolic space is by itself a well-known mathematical line of research, see for example \cite{10.2307/1999231,Tuz1992,Wang2016} and references within. As summarized in \cite{Wang2016}, a classification of minimal surfaces details the  properties of the catenoids and helicoids in  Euclidean AdS$_3$. Unfortunately, these surfaces does not necessarily correspond to  minimal surfaces dual to Wilson loops. 

Therefore, the  goal is to find minimal surfaces ending on a  boundary closed curve. This has already been achieved analytically in \cite{Ishizeki:2011bf}, where the solutions of the embedding fields are given in terms of compact Riemann surfaces of genus $g\geq 1$ \footnote{To motivate this result, recall that some nonlinear equations, such as the 2-dimensional heat equation, the KdV equation, the sine-Gordon equation among others,  are integrable and their solution are expressed in terms of Riemann theta functions.  Since theta functions can be defined from Riemann surfaces, the solutions of some nonlinear equations are related to Riemann surfaces. We refer the reader to \cite{Dubrovin1981,kalla:tel-00622289} and references within for more details. For example, in \cite{Pastras:2016vqu} they construct genus one solutions. The embedding fields are given in terms of the Weierstrass elliptic function which in turn can be written in terms of a theta function.}. It is important to stress that analytical solutions does not necessarily implies simplicity. This is the case in \cite{Ishizeki:2011bf}. The fact that the fields are given in terms of Riemann theta functions makes the problem of computing the fluctuation operator of such surface quite involved. Therefore, with the lack of the necessary computational skills, we appeal to simplicity and focus on the circular loop for the quantum correction.

\subsection{One-loop correction to the circular loop}
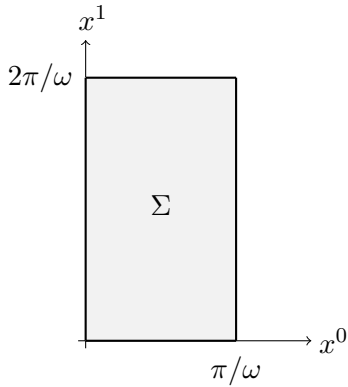
\begin{figure}[ht!]
\centering
\begin{tikzpicture}
\draw[->] (-0.1,0)--(3,0);
\draw[->] (0,-0.1)--(0,4);
\node at (3.3,0) {$x^0$};
\node at (0.1,4.3) {$x^1$};
\fill[gray!10!white] (0,0) rectangle (2,3.5);
\node at (2,-0.4) {$\pi/\omega$};
\node at (-0.6,3.5) {$2\pi/\omega$};
\node at (1,1.8){$\Sigma$};
\draw[thick] (0,0)--(2,0);
\draw[thick] (2,0)--(2,3.5);
\draw[thick] (2,3.5)--(0,3.5);
\draw[thick] (0,3.5)--(0,0);
\end{tikzpicture}
\caption{Space of parameters for the semi-sphere.}
\label{SpaceofP}
\end{figure}

It is easy to check that Eq. \eqref{semisphereparametrization1} is a solution of Eq. \eqref{eom} in the Poincar\'e half plane background
\be
  \ud s^2 =\frac{L^2}{(\phi^1)^2}\left(\left(\ud \phi^1\right)^2+\left(\ud \phi^2\right)^2+\left(\ud \phi^1\right)^3\right),
   \label{ads3metricinPC}
\ee
and take the conformally flat gauge of $h_{\mu\nu}$. The parameter space is the rectangle $\Sigma = [0,\pi/\omega]\times [0,2\pi/\omega]$, as in Figure \ref{SpaceofP}, and the circular loop is parametrized by the edge $\{0\}\times[0,2\pi/\omega] $.
In Appendix \ref{App21}, it is shown that the operator of fluctuation can be written as
\be
\cO=\frac{1}{Q^2}\tilde\cO, 
\ee
where $Q^2$ is the conformal factor of $h_{\mu\nu}$. From the quadratic action,  this factor drops out and therefore we are interested in $\tilde\cO$ 
\be
 \tilde\cO=-\left(\d^{\mu\nu}\pa_{\mu}\pa_{\nu}\mathbb{I}_{3\times 3} +\d^{\mu\nu}\upsilon_{\mu}\pa_{\nu}+\mathbb{W}\right),
\ee
where 
\bea
\upsilon_0&=&-2i\omega (\la_5 \sin(\omega x^1)+\la_2\cos(\omega x^1)),\nn
\upsilon_1&=&-2i\omega \cot(\omega x^0)(\la_2 \sin(\omega x^1)-\la_5\cos(\omega x^1)),
\eea 
and $\mathbb{W}=-Q^2(\mathbb{B}+X^R+X^{\kappa})$.
Let us consider  a  one-form connection $w_{\mu}$ and the covariant derivative
\be
\nabla_{\mu} \chi^A=\pa_{\mu} \chi^A+\d^{AB}w_{BC \mu}\chi^{C}. 
\ee
Taking $w_{\mu}=\frac{1}{2}\upsilon_{\mu}$, the operator can be written as
\be
  \tilde\cO=-(\d^{\mu\nu}\nabla_{\mu}\nabla_{\nu}\mathbb{I}+\mathbb{E}),
\ee
where
\be
 \mathbb{E}=\mathbb{W}-(\d^{\mu\nu}\pa_{\mu}w_{\nu}+\d^{\mu\nu}w_{\mu}\cdot w_{\nu}).
\ee

\begin{figure}[ht!]
\centering
\begin{tikzpicture}
\fill[gray!10!white] (0,0) rectangle (2,3.5);
\draw (0.3,0.3) rectangle (1.7,3.2);
\draw (0.3,0.3) circle (0.3cm);
\draw (1.7,0.3) circle (0.3cm);
\draw (0.3,3.2) circle (0.3cm);
\draw (1.7,3.2) circle (0.3cm);
\draw (0,0.3)--(0.3,0.3);
\draw (0.3,0.3) --(0.3,0);
\node at (0.14,-0.3){$\d/\omega$};
\draw[thick] (0,0)--(2,0);
\draw[thick] (2,0)--(2,3.5);
\draw[thick] (2,3.5)--(0,3.5);
\draw[thick] (0,3.5)--(0,0);
\end{tikzpicture}
\caption{Deformation of the space of parameters. The edges are now smooth and the circles have the same radius $\d/\omega$.}
\label{SpaceofPregulated}
\end{figure}
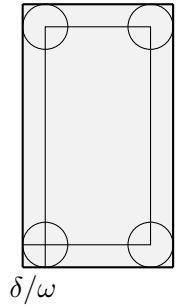

The eigenvalue problem to solve is again $\tilde\cO_{AB}\chi^B_n=\d_{AB}\la_n\chi^B_n$ with
\be
 \mu^{2}\int\ud^{2} x\,\d_{AB} \chi^A_n(x)\chi^B_m(x)=\| \vec{\chi}_n\|^2\d_{nm}.
 \label{NormEigenFunctionsOflat}
\ee
As stated in the last part of section 4, the problem can be put in a statistical mechanics auxiliary system. We consider
\be
 \d_{AB} \langle\chi^A_n|\chi^B_m\rangle =\| \vec{\chi}_n\|^2\d_{nm},\quad
\langle x| \tilde\cO_{AB}|x'\rangle = \tilde\cO_{ABx}\d(x-x').
\ee
with
\be
 1= \mu^{2}\int\ud^{2} x\, |x\rangle\langle x|,\quad\d(x-x')= \langle x|x'\rangle.
\ee
Then the partition function is given by
\be
\Tr(\ue^{-\mathscr{H}\beta})=\sum_{n}  \langle\chi^A_n|(\ue^{-\mathscr{H}\beta})_{AB}|\chi^B_n\rangle=\sum_n \| \vec{\chi}_n\|^2\ue^{-\epsilon_n\b},
\ee
where $ \mathscr{H}=L\tilde\cO$ and $\epsilon_n=L\lambda_n$. Since the modes are normalizable, we can set the norms to be one and thus obtain the partition function. It can be also written as 
\be
\cZ(\b)=  \mu^{2}\int\ud^{2} x \,\d^{AB} \langle x|(\ue^{-\mathscr{H}\beta})_{AB}|x\rangle,
\ee
where our goal is to compute the  elements $\langle x|(\ue^{-\mathscr{H}\beta})_{AB}|x\rangle$. First, we split the Hamiltonian into a free part $\mathscr{H}_0$ and interaction part $\mathscr{H}_1$
\be
 \mathscr{H}_0=-\d^{\mu\nu}\nabla_{\mu}\nabla_{\nu}\mathbb{I},\quad \mathscr{H}_1=-\mathbb{E}. 
\ee
The free Hamiltonian can be thought, like in the worldline formalism, as the quantum Hamiltonian of a non-relativistic ``coloured'' particle in the presence of non-Abelian external fields\footnote{This free Hamiltonian is known as the non-Abelian Pauli Hamiltonian which is derived from the non-relativistic limit of the Dirac equation for a
particle carrying a non-Abelian charge, see for example \cite{Anselme_Dossa_2020} and references within. Notice the absence of the electric potential in the free Hamiltonian. This implies that the particle moves in the presence of a non-Abelian external magnetic field. Also notice that the interacting Hamiltonian includes non-minimal coupling terms and interaction terms induced from the background geometry. The main difference between the hyperbolic and flat background is the absence of non-Abelian external fields for the flat case.}. 
We wish to compute the diagonal element perturbatively in $\b$ and thus one can argue on physical grounds that the first term of such an expansion should correspond to the case of the free particle decoupled to the non-Abelian external fields. We can write $\ue^{-(\pa_0^2+\pa_1^2)\beta} =\ue^{-\pa_0^2\beta} \ue^{-\pa_1^2\beta}$ and  for the one-dimensional problem on the interval $[0,T]$,  given in \cite{HKESUBEP}, we have
\begin{multline}
 \langle x|\ue^{-\pa_0^2\beta}|x'\rangle=\frac{1}{\sqrt{2\pi\b}}\left[\ue^{-\frac{(x^0-(x^0 )')^2}{4\b}}\t_3\left(\frac{(x^0-(x^0)')Ti}{2\b}\left|\frac{Ti}{\pi\b}\right.\right)	\right.\\
 \left.-\ue^{-\frac{(x^0+(x^0) ')^2}{4\b}}\t_3\left(\frac{(x^0+(x^0)')Ti}{2\b}\left|\frac{Ti}{\pi\b}\right.\right)\right],
\end{multline}
where $\t_3$ is a Jacobi theta function. At this order in the expansion, the term must be of the form $\langle x|\ue^{-\pa_0^2\beta}|x'\rangle\times \langle x|\ue^{-\pa_1^2\beta}|x'\rangle$  and we obtain
\be
\cZ(\b)\sim \mu^{2}\int\ud^{2} x \,\d^{AB} \frac{1}{4\pi\b}\d_{AB }= \mu^2 \frac{3\mathrm{Vol}(\Sigma)}{4\pi\b}.
\ee
In general, we write the asymptotic expansion of the partition function as 
\be
 \cZ(\b)\sim \mu^2\b^{-(1+p)/2}\sum_{l=0}\b^{l/2}a_l.
\ee
This is the Seeley-DeWitt expansion and again as stated in \cite{BransonGilkey90,Vassilevich:2003xt}, 
the half integer powers in $\b$ are due to the presence of boundaries. The first coefficient is $a_0 = (3\mathrm{Area}(\Sigma))/(4\pi)=(3\pi)/(2\omega^2)$. The terms $a_{1}$ and $a_2$ are given in   \cite{BransonGilkey90,Vassilevich:2003xt}. For our problem, they correspond to 
\bea
 a_1&=&-\frac{1}{4}\frac{1}{\sqrt{4\pi}}3\times \mathrm{Perimeter}(\Sigma)=-\frac{9}{4}\frac{\sqrt{\pi}}{\omega},\\
  a_2&=& \frac{1}{4\pi} \int\ud^{2} x \,\d^{AB} \mathbb{E}_{AB},\nn
  &=&\frac{1}{4\pi} \int\ud^{2} x \left[\frac{4\omega^2}{\sin^2(\omega x^0)}+(4\omega^4L^2-\omega^2)\cot^2(\omega x^0)+4\omega^4L^2\left(\cot^4(\omega x^0)+\frac{1}{\sin^4(\omega x^0)}\right)\right],\nn
  &=&\frac{\pi}{2}-\left(\frac{3}{2}+2\omega^2L^2\right)\left.(\cot(u))\right|_0^{\pi}-\frac{4}{3}\omega^2L^2\left.(\cot^3(u))\right|_0^{\pi}.
\eea
We note that $a_2$ diverges and the endpoints of $\Sigma$ are on the $x^0$-axis. We regulate $\Sigma$ by considering the smoothing of the edges by the deformation $\Sigma'$ shown in Figure \ref{SpaceofPregulated}.
The regulated area and perimeter are
\be
 \mathrm{Area}(\Sigma')=\mathrm{Area}(\Sigma)-(4-\pi)\frac{\d^2}{\omega^2},\quad \mathrm{Perimeter}(\Sigma')=\mathrm{Perimeter}(\Sigma)-2(4-\pi)\frac{\d}{\omega},
\ee
and the regulated coefficients result 
\bea
a'_0&=&\frac{3\pi}{2\omega^2}-\frac{3}{4\pi}(4-\pi)\frac{\d^2}{\omega^2},\nn
a'_1&=&-\frac{9}{4}\frac{\sqrt{\pi}}{\omega}+\frac{3}{4\sqrt{\pi}}(4-\pi)\frac{\d}{\omega},\nn
a'_2&\approx &\frac{\pi}{2}+\left(\frac{9}{4}+3\omega^2L^2\right)\frac{1}{\d}+\frac{3}{2}\omega^2L^2\frac{1}{\d^3}.
\label{regcoeffa012}
\eea
Hence, the renormalized partition function is
\be
\frac{\cZ(\b)^{\mathrm{renorm.}}}{\mu^2}\sim  \frac{3\pi}{2\omega^2}\frac{1}{\beta} -\frac{9}{4}\frac{\sqrt{\pi}}{\omega}\frac{1}{\sqrt{\b}}+\frac{\pi}{2}+O(\sqrt{\b}).
\ee
Let us write then the zeta function in terms of $\b$
\be
\zeta(s,\tilde{\cO})=\frac{\mu^2}{L^s}\frac{1}{\Gamma(s)}\int\limits_{\b_H}^{\b_L}\ud \b\,\b^{s-1}\frac{\cZ(\b)}{\mu^2},
\ee
where $\b_h$ and $\b_L$ corresponds to high and low inverse temperature cutoffs respectively. Then
\be
\ue^{-S_{(1)}}\Leftrightarrow \frac{1}{\sqrt{\det\tilde\cO'}}=\ue^{-\frac{\mu^2}{2}\left(-\frac{3\pi}{2\omega^2}\frac{1}{\b_H}+\frac{9\sqrt{\pi}}{4\omega}\frac{2}{\sqrt{\b_H}}-\frac{\pi}{2}\ln\left(\frac{\b_L}{\b_H}\right)\right)}.\label{HeatOneloopEffectiveaction}
\ee
We see that a $\cS_1\to \infty$ divergence correspond to the limit $\b_H\to 0$ in the perimeter term, i.e. the term proportional to the inverse of $\omega$. This term should correspond to the term given in Eq.\eqref{HRoneloop} in the canonical analysis and we can write symbolically the relation between the regularization schemes as 
\be
\frac{\overline{\mathrm{Perimeter}(\Sigma')}}{\sqrt{\b_H}}\Leftrightarrow \ln \left(\frac{\mathrm{Perimeter\,of\,the\,circle}}{\epsilon}\right),
\ee
for $\b_H,\epsilon\to 0$. The expression $\overline{\mathrm{Perimeter}(\Sigma')}$ indicates that we only consider the vertical curve close to $x^0=0$ of the perimeter of $\Sigma'$ since it corresponds to the subspace that describes the circle. For $\b_H\to 0$ or $\b_L\to\infty$, the remaining terms in Eq. \eqref{HeatOneloopEffectiveaction} break the validity of the WKB approximation. 

Lastly, using \eqref{densityrho(u)} with $v^2\approx u$ at high temperatures and the rescalings $a_0\to a_0/L$, $a_1\to a_1/\sqrt{L}$, the density is
\be
\rho(v')\approx\frac{3\pi}{\omega^2}\mu^2 v' -\frac{9}{2}\frac{\mu^2}{\omega} +\pi\mu^2 v'\d((v')^2).
\ee
The first two terms correspond to the free and boundary part of the Seeley-DeWitt expansion respectively. This in agreement with the calculation of the counting function of the minus Laplacian in a rectangle performed in \cite{Dai:2009zza} after taking into account the ``colour'' degrees of freedom.  The third term must be  related to the interactions and indeed this is the case since the $a_2'$ coefficient depends on $\mathscr{H}_1$. We should expect that this term contributes to the phase shift but as discussed for the worldline,  using a computation analogous to Eq. \eqref{diracandphase}, we shown that there is no such contribution at this order.

\subsection{Zero modes}
Using \eqref{zeromodesformula} and \eqref{semisphereparametrization1} we find the zero modes are
\be
\vec{\chi}_0^{(1)}=\begin{pmatrix}
\sin(\omega x^0)\\
\cos(\omega x^1)\cos(\omega x^0)\\
\sin(\omega x^1)\cos(\omega x^0)
\end{pmatrix},\quad
\vec{\chi}_0^{(2)}=\begin{pmatrix}
0\\
1\\
0
\end{pmatrix},\quad
\vec{\chi}_0^{(3)}=\begin{pmatrix}
0\\
0\\
1
\end{pmatrix},
\label{ZEROMODESADS3}
\ee
related to dilatations and translational invariance respectively. In order to write the wavefunction as Eq. \eqref{zero1loopintegral2}, we first compute the on-shell action
\bea
S^*&=&M_{1}\frac{2\pi}{\omega} \int\limits_0^{\frac{\pi}{\omega}}\ud x^0\, \frac{L^2\omega^2}{\sin^2(\omega x^0)}\cos(\omega x^0),\nn
&=&M_{1}\frac{2\pi}{\omega} (L^2\omega\rho_0)\int\limits_{\epsilon}^{\rho_0}\ud \phi^1 \frac{1}{(\phi^1)^2},\nn
&=&-M_1L^2 2\pi+M_1L\left(\frac{L}{\epsilon}\right)2\pi\rho_0.
\label{actiononshellCWL}
\eea
Notice that the first term is independent of $\omega$ and $\rho$. The second term agrees with the classical part of the Wheeler-DeWitt solution given by Eq. \eqref{perimeterlaw} after identifying $\mu= 1/L$. Then the wavefunction is written as 
\be 
 \Psi \approx \cN' \int \ud\rho_0\ud\phi^2_0\ud\phi^3_0\,\left(\frac{\sqrt{\pi M_{(1)}}}{\omega}\right)^3\,\frac{\ue^{-S^{*}}}{\sqrt{\det(\cO'/\mu^2)}}
 =\cN''\left(\frac{1}{\sqrt{2}\omega}\right)^3\la^{3/4}\frac{\ue^{-S^{*}}}{\sqrt{\det(\cO'/\mu^2)}},
 \label{zero1loopintegralcircularWL}
\ee
where in the second equality, we  introduced the t'Hooft coupling $\la$ defined as $M_1\equiv\sqrt{\la}/2\pi$. The integration over the moduli restored some target isometries from  spontaneously breaking. For completness, let us review the classification of the target isometries in our problem. It is well known  that the 3-dimensional hyperbolic manifold orientation-preserving isometry group is $\mathrm{PSL}(2,\mathbb{C})$. The plane $\phi^1=0$ in the upper-half model, can be seen as the complex plane $\mathbb{C}$ with the addition of the point `$\infty$'. We conclude that $\hat{\mathbb{C}}=\mathbb{C} \cup \{\infty\}$, i.e. the Riemann sphere, can be thought as the boundary of hyperbolic 3-space. Every M\"obius transformation maps the Riemann sphere to itself. The key point to notice is that every orientation-preserving isometry of the hyperbolic 3-space can be seen as an extensions of  M\"obius transformations on the Riemann sphere. Following \cite{benedetti1992lectures}, the isometries that have no fixed point in the hyperbolic 3-space are of the form
\be
 (\phi^1,X)\to (\phi^1,X+\la),\quad (\phi^1,X)\to (|\la|\phi^1,\la X),
\ee
where $\la \in \mathbb{C}$ and $X=\phi^2+i\phi^3$. The classification of the isometries follows from the  four types of non-identity elements of $\mathrm{PSL}(2,\mathbb{C})$: elliptic, parabolic and hyperbolic. The case $ (\phi^1,X)\to (\phi^1,X+\la)$ corresponds to a parabolic type (its action has two fixed points at $\infty$ in $\hat{\mathbb{C}}$ but not fixed point in the hyperbolic 3-space). The case $(\phi^1,X)\to (|\la|\phi^1,\la X)$ with $|\la|\neq 1$ corresponds to the hyperbolic type (both $\infty$ and $0$ are fixed points in $\hat{\mathbb{C}}$ but not fixed point in the hyperbolic 3-space). Therefore we can classify the zero modes in our problem as parabolic and hyperbolic. Regarding the remaining isometry,  the semi-sphere minimal surface $(\phi^1)^2+X\bar{X}=1$ is invariant under the elliptic type ($|\la|=1$, i.e. rotations in the complex plane) and thus there is no spontaneous symmetry breaking for this isometry.
 
We conclude this section by adding an interaction term to the Dirac action and consider to the one-loop correction the sum over the topologies. Let us assume that the total action is now given by the Dirac action plus an interaction term proportional to the Euler number of the 2-brane. As discussed in \cite{Drukker:2000rr}, the interaction will multiply the wavefunctional with a factor $\la^{\frac{6p-3}{4}}$, where $p$ is the number of handles. The number $6p-3$ corresponds to the dimension of the moduli space of surfaces of genus $p$ with one boundary. For genus 0, the factor is $\la^{-\frac{3}{4}}$ and thus cancelling the zero modes contribution.

\section{Conclusions and discussion}
The one-loop correction to the wavefunctional of a bosonic extended objects in a hyperbolic background has been studied via the path integral and the Wheeler-DeWitt equation. We have found a semiclassical solution of the Wheeler-DeWitt equation that provides an identification of divergences at the classical and quantum level. This in the same vein of the holographic renormalization program and the one-loop results of our solution gives a novel extension. From the path integral perspective, the one-loop functional determinant has been regulated by the heat kernel method. The  high energy/temperature of the heat kernel provides in a systematic fashion the expected UV divergences. Therefore we can set up a framework to deal with the divergences with the aid of an auxiliary quantum statistical mechanics system, as done in \cite{Mukhanov:2007zz} and in the worldline formalism.

We have seen that a primary classification of these divergences follows from the compatibility with the semiclassical approximation. More precisely, the one-loop correction can either vanish or blow up independently of the classical contribution. For the second case, the WKB approximation breaks down. A secondary classification shows the geometrical origin of such divergences. This can be done since the coefficients of the leading divergences in the high energy/temperature of the heat kernel have a geometrical meaning; they correspond to volume and boundary of the space of parameters $\Sigma$.  

For the circular Wilson loop in Euclidean AdS$_3$ discussed in section 5, the constant term in the heat kernel expansion gives a logarithmic divergence that breaks the WKB approximation. This behaviour is expected to be true for any dimension unless the coefficient becomes negative. The divergence related to  the perimeter $\pa\Sigma$ does not break the WKB approximation. Moreover, due to the semiclassical solution of the Wheeler-DeWitt equation, we learned that this divergence corresponds to the one-loop divergence of the perimeter of the boundary curve in the hyperbolic space and this provides a relation between regulating schemes.

The relation with the heat kernel expansion and the phase shift has been discussed. We have shown that only subleading terms in the high energy/temperature contributes to the semiclassical expression of the phase shift. We also clarify that a naive computation of the  phase shift from  a term in the expansion that gives a Dirac delta term in the density of states. However, the solution of the Wheeler-DeWitt equation tells us that this consideration is physically incorrect.

Having a clear picture of the one-loop correction, we can extend the result of the circular loop in Euclidean AdS$_3$  to  an arbitrary closed curve in the boundary as developed in \cite{Ishizeki:2011bf,Kruczenski:2013bsa,Kruczenski:2014bla,Irrgang:2015txa,Huang:2016atz,He:2017zsk,He:2017cwd,Cooke:2018obg}. Let the parameter space $\Sigma$ be a smooth region. The loop in the boundary of Euclidean AdS$_3$ is parametrized by 
$\pa\Sigma$ or a part of it. As stated in \cite{Nesterenko:2002ng}, there are new curvature contributions to the trace of the heat kernel for the operator $-\d^{\mu}\pa_{\mu}\pa_{\nu}$. Assuming  Dirichlet boundary conditions on $\pa\Sigma$, the contributions are of the form
\be
K(\tau)=\frac{\Tr\mathrm{Vol}(\Sigma)}{4\pi\tau}- \frac{\Tr\mathrm{Vol}(\pa\Sigma)}{4\sqrt{4\pi \tau}} +\frac{1}{12\pi}\int_{\pa\Sigma}\ud s\, \kappa(s) +O(\sqrt{\tau},\kappa^2),
\ee
where $\kappa$ is the curvature of the closed curve and $\Tr$ is taken over the ``coloured'' degrees of freedom. For a general operator, the interaction terms are subleading in the expansion and we just need to add to  the above expression the $\tau$-independent contribution of $a_2$.

Finally, we briefly comment on the possible cancellation of divergences. Clearly, if the underling theory is conformally invariant, i.e. our starting point is the Polyakov action instead of the Dirac action,  there must be an internal cancellation of these scales provided by the degrees of freedom in the theory and the same reasoning applies for a supersymmetric theory. Examples of the circuitry of such cancellations has been discussed explicitly for holographic Wilson loops in \cite{Faraggi:2016ekd,David:2019lhr}. This will be studied, for models such as the ones developed here, in a future work.

\subsection*{Acknowledgements}
I have benefited from several discussions with Marina David, Alberto Faraggi,  Leopoldo Pando Zayas and Guillermo Silva on the subjects of holographic Wilson loops and regularization methods. I would like to thank the anonymous reviewers for their comments and suggestions that improved and clarified this manuscript. This work was funded by the National Agency for Research and Development (ANID), Concurso FONDECYT de Postdoctorado 2020 \# 3200721.

\appendix
\section{Geometrical data of the worldline in $d=2$}
\label{App1}
In two dimensions, the ambient metric is of the form
\be
g=f^2(\phi^1,\phi^2)\left[(\ud \phi^1)^2+(\ud \phi^2)^2\right]. 
\ee
Then we have $e^A=\d^A_{\;\;\; i}f\ud \phi^i$. In components
\be
 e^1_{\;\;\; 1}=f,\quad e^1_{\;\;\; 2}=0, \quad e^2_{\;\;\; 1}=0,\quad e^2_{\;\;\; 2}=f.
\ee
Since the one-from connection is the Levi-Civita spin connection, we have 
\be
\Omega_{AB}=-\Omega_{BA}, \quad \ud e^A +\Omega^A_{\;\;\;B}\wedge e^B=0.
\label{Cartan1}
\ee
Therefore, we must solve 
\be
\ud e^1+ \Omega^1_{\;\;\;2}\wedge e^2=0,\quad \ud e^2 +\Omega^2_{\;\;\;1}\wedge e^1=0.
\label{Cartan1d2}
\ee
From
\bea
\ud  e^1 &=& \pa_if \ud\phi^i \wedge\ud \phi^1 =  \pa_2 f \ud\phi^2 \wedge\ud \phi^1=\frac{1}{f}\pa_2\ln f\, e^2\wedge e^1,\\
\ud  e^2 &=& \pa_if \ud\phi^i \wedge\ud \phi^2 =  \pa_1 f \ud\phi^1 \wedge\ud \phi^2=\frac{1}{f}\pa_1\ln f\, e^1\wedge e^2,
\eea
together with the ansatz $\Omega^1_{\;\;\;2}=\a e^1+\b e^2$ and Eq. \eqref{Cartan1d2}, we obtain 
\be
 \Omega^1_{\;\;\;2}= \pa_2\ln f\,\ud \phi^1-\pa_1\ln f\,\ud \phi^2,
\ee
The second equation to satisfy is 
\be
\ud \Omega^A_{\;\;\;B} +\Omega^A_{\;\;\;C}\wedge \Omega^C_{\;\;\;B}= R^A_{\;\;\;B} = \frac{1}{2}R^A_{\;\;\;BCD} \,e^C\wedge e^D 
\label{Cartan2}
\ee
Then 
\be
 R^1_{\;\;\;2}= \ud \Omega^1_{\;\;\;2}=-\Box \ln f\, \ud\phi^1\wedge\ud\phi^2, \quad  \Box=\d^{AB}\pa_A\pa_B,
\ee
and
\be
R^1_{\;\;\;212}=-\frac{1}{f^2}\Box \ln f.
\ee

\section{Geometrical data of Euclidean AdS$_3$}
\label{App21}
Ihe ambient metric is given in Eq. \eqref{ads3metricinPC} and we have
\be
e^A=\frac{L}{\phi^1}\d^A_{\;\;\; i}\ud \phi^i, \quad \ud e^A=\frac{1}{L}e^A\wedge e^1.
\ee
The solution of Eq. \eqref{Cartan1} is
\be
 \Omega^1_{\;\;\;2}=\frac{1}{L}e^2,\quad \Omega^1_{\;\;\;3}=\frac{1}{L}e^3,\quad \Omega^2_{\;\;\;3}=0.
\ee
From Eq. \eqref{Cartan2} we obtain
\be
 R^1_{\;\;\;2}=\frac{1}{L^2}e^2\wedge e^1, \quad  R^1_{\;\;\;3}=\frac{1}{L^2}e^3\wedge e^1,  R^2_{\;\;\;3}=\frac{1}{L^2}e^3\wedge e^2,
\ee
and the non-zero components 
\be 
R^1_{\;\;212} =-\frac{1}{L^2},\quad R^1_{\;\;313} =-\frac{1}{L^2},\quad R^2_{\;\;323} =-\frac{1}{L^2}.
\ee
\subsection{Data from the solution}
\label{App22}
The equations of motion for $h_{\mu\nu}=Q^2\d_{\mu\nu}$ are
\be
\d^{\mu\nu}\pa_{\mu}\pa_{\nu}\phi^i+\Gamma^i_{lm}\d^{\mu\nu}\pa_{\mu}\phi^l\pa_{\nu}\phi^m=\d^{\mu\nu}\kappa^i_{\mu\nu}=0,
\ee
and the non-vanishing Christoffel symbols are
\be
 \Gamma^1_{11}=-\frac{1}{\phi^1},\quad  \Gamma^1_{22}=\frac{1}{\phi^1},\quad \Gamma^1_{33}=\frac{1}{\phi^1},\quad \Gamma^2_{12}=-\frac{1}{\phi^1},\quad \Gamma^3_{13}=-\frac{1}{\phi^1}.
\ee
Then
\bea
k^1_{\mu\nu}&=&\begin{pmatrix}
-\rho_0\omega^2\cot(\omega x^0)\cos(\omega x^0)&0\\
0&\rho_0\omega^2\cot(\omega x^0)\cos(\omega x^0)
\end{pmatrix},\nn
&=&-\rho_0\omega^2\cot(\omega x^0)\cos(\omega x^0)\sigma_3,\nn
k^2_{\mu\nu}&=&\begin{pmatrix}
\rho_0\omega^2\cos(\omega x^1)\cos(\omega x^0)&\rho_0\omega^2\frac{\sin(\omega x^1)}{\sin(\omega x^0)}\\
\rho_0\omega^2\frac{\sin(\omega x^1)}{\sin(\omega x^0)}&-\rho_0\omega^2\cos(\omega x^1)\cos(\omega x^0)\end{pmatrix},\nn
&=&\rho_0\omega^2\cos(\omega x^1)\cos(\omega x^0)\sigma_3+\rho_0\omega^2\frac{\sin(\omega x^1)}{\sin(\omega x^0)}\sigma_1,\nn
k^3_{\mu\nu}&=&\begin{pmatrix}
\rho_0\omega^2\sin(\omega x^1)\cos(\omega x^0)&-\rho_0\omega^2\frac{\cos(\omega x^1)}{\sin(\omega x^0)}\\
-\rho_0\omega^2\frac{\cos(\omega x^1)}{\sin(\omega x^0)}&-\rho_0\omega^2\sin(\omega x^1)\cos(\omega x^0)\end{pmatrix},\nn
&=&\rho_0\omega^2\sin(\omega x^1)\cos(\omega x^0)\sigma_3-\rho_0\omega^2\frac{\cos(\omega x^1)}{\sin(\omega x^0)}\sigma_1.
\label{kappamatrices}
\eea
The components of the pullback spin connection are 
\bea 
\omega_{12\mu}&=&\begin{pmatrix}
-\omega \cos(\omega x^1),\\
-\omega \sin(\omega x^1)\cot(\omega x^0)
\end{pmatrix},\nn
\omega_{13\mu}&=&\begin{pmatrix}
-\omega \sin(\omega x^1),\\
\omega \cos(\omega x^1)\cot(\omega x^0).
\end{pmatrix}.
\eea
The antisymmetric part of the operator of the fluctuation is
\begin{multline}
\mathbb{A}=\frac{1}{Q^2}\left[i\la_22\omega(\cos(\omega x^1)D_0+\sin(\omega x^1)\cot(\omega x^0)D_1)\right.\\
\left.+i\la_5 2\omega(\sin(\omega x^1)D_0-\cos(\omega x^1)\cot(\omega x^0)D_1)\right],
\end{multline}
where the $\lambda$ matrices are the Gell-Mann matrices
\be
 \la_2=\begin{pmatrix}
 0&-i&0\\
 i&0&0\\
 0&0&0
 \end{pmatrix},\quad 
  \la_5=\begin{pmatrix}
 0&0&-i\\
 0&0&0\\
 i&0&0
 \end{pmatrix}.
\ee
The term proportional to $(\omega_{AB\mu})^2$ of the operator is given by
\be
\mathbb{B}=\frac{\omega^2}{Q^2}\begin{pmatrix}
\frac{1}{\sin^2(\omega x^0)}&0&0\\
0&\cos^2(\omega x^1)+\sin^2(\omega x^1)\cot^2(\omega x^0)&\cos(\omega x^1)\sin(\omega x^1)(1-\cot^2(\omega x^0))\\
0&\cos(\omega x^1)\sin(\omega x^1)(1-\cot^2(\omega x^0))&\sin^2(\omega x^1)+\cos^2(\omega x^1)\cot^2(\omega x^0)
\end{pmatrix} .
\ee

The symmetric matrix $X_{AB}$ can be decompose into two parts: $X_{AB}^R$, which depends on the Riemann tensor and $X_{AB}^{\kappa}$, which depends on $\kappa_{\mu\nu}^2$. Direct computation gives
\bea
 X_{11}^R&=&\frac{\omega^2}{Q^2}\frac{1}{\sin^2(\omega x^0)},\nn
 X_{12}^R&=&\frac{\omega^2}{Q^2}\cos(\omega x^1)\cot(\omega x^0),\nn
 X_{13}^R&=&\frac{\omega^2}{Q^2}\sin(\omega x^1)\cot(\omega x^0),\nn
 X_{22}^R&=&\frac{\omega^2}{Q^2}\left[\sin^2(\omega x^1)+\cot^2(\omega x^0)(\cos^2(\omega x^1)+1)\right],\nn
X_{23}^R&=&\frac{\omega^2}{Q^2}\cos(\omega x^1)\sin(\omega x^1)(\cot^2(\omega x^0)-1),\nn
X_{33}^R&=&\frac{\omega^2}{Q^2}\left[\cos^2(\omega x^1)+\cot^2(\omega x^0)(\sin^2(\omega x^1)+1)\right].
\eea
For the matrix $X_{AB}^{\kappa}$, we see that it can be written in a matrix notation as
\be
X_{AB}^{\kappa}=-2\frac{1}{Q^2}\frac{L^2}{(\phi^1)^2}\d_{AC}\d_{BD}\d^{\mu\nu}(\kappa^C\kappa^D)_{\mu\nu},
\ee
where the matrices $\kappa^A$ are given in Eq. \eqref{kappamatrices}. Then
\bea
X_{11}^{\kappa} &=& -4\frac{\omega^4L^2}{Q^2}\cot^4(\omega x^0),\nn
X_{12}^{\kappa} &=&4\frac{\omega^4L^2}{Q^2}\cot^3(\omega x^0)\cos(\omega x^1),\nn
X_{13}^{\kappa} &=&4\frac{\omega^4L^2}{Q^2}\cot^3(\omega x^0)\sin(\omega x^1),\nn
X_{22}^{\kappa} &=&-4\frac{\omega^4L^2}{Q^2}\left(\cos^2(\omega x^1)\cot^2(\omega x^0)+\frac{\sin^2(\omega x^1)}{\sin^4(\omega x^0)}\right),\nn
X_{23}^{\kappa} &=&-4\frac{\omega^4L^2}{Q^2}\cos(\omega x^1)\sin(\omega x^1)\left(\cot^2(\omega x^0)-\frac{1}{\sin^4(\omega x^0)}\right),\nn
X_{33}^{\kappa} &=&-4\frac{\omega^4L^2}{Q^2}\left(\sin^2(\omega x^1)\cot^2(\omega x^0)+\frac{\cos^2(\omega x^1)}{\sin^4(\omega x^0)}\right).
\eea
Hence, the operator can be written as
\be
\cO=\frac{1}{Q^2}\tilde\cO, 
\ee
where 
\begin{multline}
 \tilde\cO=-(\pa_0^2+\pa_1^2)\mathbb{I} +i\la_22\omega(\cos(\omega x^1)D_0+\sin(\omega x^1)\cot(\omega x^0)D_1)\\
+i\la_5 2\omega(\sin(\omega x^1)D_0-\cos(\omega x^1)\cot(\omega x^0)D_1)+\mathbb{B}+X^R+X^{\kappa}.
\end{multline}

\bibliographystyle{unsrt}
\bibliography{Sigmamodels} 

\begin{thebibliography}{100}

\bibitem{Maldacena:1997re}
Juan~Martin Maldacena.
\newblock {The Large N limit of superconformal field theories and
  supergravity}.
\newblock {\em Int. J. Theor. Phys.}, 38:1113--1133, 1999.

\bibitem{Kinar:1999xu}
Y.~Kinar, E.~Schreiber, J.~Sonnenschein, and N.~Weiss.
\newblock {Quantum fluctuations of Wilson loops from string models}.
\newblock {\em Nucl. Phys.}, B583:76--104, 2000.

\bibitem{Forste:1999qn}
Stefan Forste, Debashis Ghoshal, and Stefan Theisen.
\newblock {Stringy corrections to the Wilson loop in N=4 superYang-Mills
  theory}.
\newblock {\em JHEP}, 08:013, 1999.

\bibitem{Drukker:2000ep}
Nadav Drukker, David~J. Gross, and Arkady~A. Tseytlin.
\newblock {Green-Schwarz string in AdS$_5 \times$ S$^5$: Semiclassical
  partition function}.
\newblock {\em JHEP}, 04:021, 2000.

\bibitem{Kruczenski:2008zk}
M.~Kruczenski and A.~Tirziu.
\newblock {Matching the circular Wilson loop with dual open string solution at
  1-loop in strong coupling}.
\newblock {\em JHEP}, 05:064, 2008.

\bibitem{Beccaria:2010zn}
M.~Beccaria, G.~V. Dunne, G.~Macorini, A.~Tirziu, and A.~A. Tseytlin.
\newblock {Exact computation of one-loop correction to energy of pulsating
  strings in AdS$_5 \times$ S$^5$}.
\newblock {\em J. Phys.}, A44:015404, 2011.

\bibitem{Faraggi:2011bb}
Alberto Faraggi and Leopoldo~A. Pando~Zayas.
\newblock {The Spectrum of Excitations of Holographic Wilson Loops}.
\newblock {\em JHEP}, 05:018, 2011.

\bibitem{Faraggi:2011ge}
Alberto Faraggi, Wolfgang Mueck, and Leopoldo~A. Pando~Zayas.
\newblock {One-loop Effective Action of the Holographic Antisymmetric Wilson
  Loop}.
\newblock {\em Phys. Rev.}, D85:106015, 2012.

\bibitem{Kristjansen:2012nz}
Charlotte Kristjansen and Yuri Makeenko.
\newblock {More about One-Loop Effective Action of Open Superstring in
  $AdS_5\times S^5$}.
\newblock {\em JHEP}, 09:053, 2012.

\bibitem{Kim:2012tu}
Hyojoong Kim, Nakwoo Kim, and Jung Hun~Lee.
\newblock {One-loop corrections to holographic Wilson loop in AdS4xCP3}.
\newblock {\em J. Korean Phys. Soc.}, 61:713--719, 2012.

\bibitem{Forini:2014kza}
Valentina Forini, Valentina Giangreco~M. Puletti, Michael Pawellek, and Edoardo
  Vescovi.
\newblock {One-loop spectroscopy of semiclassically quantized strings: bosonic
  sector}.
\newblock {\em J. Phys.}, A48(8):085401, 2015.

\bibitem{Buchbinder:2014nia}
E.~I. Buchbinder and A.~A. Tseytlin.
\newblock {1/N correction in the D3-brane description of a circular Wilson loop
  at strong coupling}.
\newblock {\em Phys. Rev.}, D89(12):126008, 2014.

\bibitem{Forini:2015mca}
V.~Forini, V.~Giangreco~M. Puletti, L.~Griguolo, D.~Seminara, and E.~Vescovi.
\newblock {Remarks on the geometrical properties of semiclassically quantized
  strings}.
\newblock {\em J. Phys.}, A48(47):475401, 2015.

\bibitem{Bergamin:2015vxa}
R.~Bergamin and A.~A. Tseytlin.
\newblock {Heat kernels on cone of $AdS_2$ and $k$-wound circular Wilson loop
  in $AdS_5 \times S^5$ superstring}.
\newblock {\em J. Phys.}, A49(14):14LT01, 2016.

\bibitem{Faraggi:2016ekd}
Alberto Faraggi, Leopoldo~A. Pando~Zayas, Guillermo~A. Silva, and Diego
  Trancanelli.
\newblock {Toward precision holography with supersymmetric Wilson loops}.
\newblock {\em JHEP}, 04:053, 2016.

\bibitem{Forini:2017whz}
V.~Forini, A.~A. Tseytlin, and E.~Vescovi.
\newblock {Perturbative computation of string one-loop corrections to Wilson
  loop minimal surfaces in AdS$_5 \times$ S$^5$}.
\newblock {\em JHEP}, 03:003, 2017.

\bibitem{Cagnazzo:2017sny}
Alessandra Cagnazzo, Daniel Medina-Rincon, and Konstantin Zarembo.
\newblock {String corrections to circular Wilson loop and anomalies}.
\newblock {\em JHEP}, 02:120, 2018.

\bibitem{Chen-Lin:2017pay}
Xinyi Chen-Lin, Daniel Medina-Rincon, and Konstantin Zarembo.
\newblock {Quantum String Test of Nonconformal Holography}.
\newblock {\em JHEP}, 04:095, 2017.

\bibitem{Medina-Rincon:2018wjs}
Daniel Medina-Rincon, Arkady~A. Tseytlin, and Konstantin Zarembo.
\newblock {Precision matching of circular Wilson loops and strings in AdS$_{5}$
  × S$^{5}$}.
\newblock {\em JHEP}, 05:199, 2018.

\bibitem{Aguilera-Damia:2018bam}
Jeremías Aguilera-Damia, Alberto Faraggi, Leopoldo~A. Pando~Zayas, Vimal
  Rathee, and Guillermo~A. Silva.
\newblock {Toward Precision Holography in Type IIA with Wilson Loops}.
\newblock {\em JHEP}, 08:044, 2018.

\bibitem{Medina-Rincon:2019bcc}
Daniel Medina-Rincon.
\newblock {Matching quantum string corrections and circular Wilson loops in
  $AdS_4 \times CP^3$}.
\newblock {\em JHEP}, 08:158, 2019.

\bibitem{David:2019lhr}
Marina David, Rodrigo de~Le\'on~Ard\'on, Alberto Faraggi, Leopoldo~A.
  Pando~Zayas, and Guillermo~A. Silva.
\newblock {One-loop holography with strings in $AdS_4\times\mathbb {CP}^3$}.
\newblock {\em JHEP}, 10:070, 2019.

\bibitem{Hernandez:2019huf}
Rafael Hernández, Juan~Miguel Nieto, and Roberto Ruiz.
\newblock {Quantum corrections to minimal surfaces with mixed three-form flux}.
\newblock {\em Phys. Rev.}, D101(2):026019, 2020.

\bibitem{Henneaux:1992ig}
M.~Henneaux and C.~Teitelboim.
\newblock {\em {Quantization of gauge systems}}.
\newblock 1992.

\bibitem{Halliwell:1990uy}
Jonathan~J. Halliwell.
\newblock {Introductory lectures on quantum cosmology}.
\newblock In {\em {7th Jerusalem Winter School for Theoretical Physics: Quantum
  Cosmology and Baby Universes}}, pages 159--243, 1989.

\bibitem{Hawking:1980gf}
S.W. Hawking.
\newblock {\em {The path integral approach to quantum gravity}}, pages
  746--789.
\newblock 1 1980.

\bibitem{Vilenkin:1982de}
Alexander Vilenkin.
\newblock {Creation of Universes from Nothing}.
\newblock {\em Phys. Lett. B}, 117:25--28, 1982.

\bibitem{Hartle:1983ai}
J.B. Hartle and S.W. Hawking.
\newblock {Wave Function of the Universe}.
\newblock {\em Adv. Ser. Astrophys. Cosmol.}, 3:174--189, 1987.

\bibitem{Vilenkin:1986cy}
Alexander Vilenkin.
\newblock {Boundary Conditions in Quantum Cosmology}.
\newblock {\em Phys. Rev. D}, 33:3560, 1986.

\bibitem{Vilenkin:1987kf}
Alexander Vilenkin.
\newblock {Quantum Cosmology and the Initial State of the Universe}.
\newblock {\em Phys. Rev. D}, 37:888, 1988.

\bibitem{Halliwell:1988wc}
Jonathan~J. Halliwell.
\newblock {Derivation of the Wheeler-De Witt Equation from a Path Integral for
  Minisuperspace Models}.
\newblock {\em Phys. Rev. D}, 38:2468, 1988.

\bibitem{Halliwell:1989dy}
Jonathan~J. Halliwell and James~B. Hartle.
\newblock {Integration Contours for the No Boundary Wave Function of the
  Universe}.
\newblock {\em Phys. Rev. D}, 41:1815, 1990.

\bibitem{Voros:1986vw}
A.~Voros.
\newblock {Spectral Functions, Special Functions and Selberg Zeta Function}.
\newblock {\em Commun. Math. Phys.}, 110:439, 1987.

\bibitem{voros1992}
André Voros.
\newblock Spectral zeta functions.
\newblock In {\em Zeta Functions in Geometry}, pages 327--358, Tokyo, Japan,
  1992. Mathematical Society of Japan.

\bibitem{jorgenson1993basic}
J.~Jorgenson and S.~Lang.
\newblock {\em Basic analysis of regularized series and products}.
\newblock Lecture notes in mathematics. Springer-Verlag, 1993.

\bibitem{Vassilevich:2003xt}
D.V. Vassilevich.
\newblock {Heat kernel expansion: User's manual}.
\newblock {\em Phys. Rept.}, 388:279--360, 2003.

\bibitem{Fursaev:2011zz}
Dmitri Fursaev and Dmitri Vassilevich.
\newblock {\em {Operators, Geometry and Quanta}}.
\newblock Theoretical and Mathematical Physics. Springer, Berlin, Germany,
  2011.

\bibitem{Forini:2017ene}
Valentina Forini.
\newblock {On regulating the AdS superstring}.
\newblock pages 221--244. 2018.

\bibitem{Aguilera-Damia:2018rjb}
Jeremías Aguilera-Damia, Alberto Faraggi, Leopoldo Pando~Zayas, Vimal Rathee,
  and Guillermo~A. Silva.
\newblock {Functional Determinants of Radial Operators in $AdS_2$}.
\newblock {\em JHEP}, 06:007, 2018.

\bibitem{Aguilera-Damia:2018twq}
Jeremías Aguilera-Damia, Alberto Faraggi, Leopoldo~A. Pando~Zayas, Vimal
  Rathee, and Guillermo~A. Silva.
\newblock {Zeta-function Regularization of Holographic Wilson Loops}.
\newblock {\em Phys. Rev.}, D98(4):046011, 2018.

\bibitem{Gervais:1974dc}
Jean-Loup Gervais and B.~Sakita.
\newblock {Extended Particles in Quantum Field Theories}.
\newblock {\em Phys. Rev. D}, 11:2943, 1975.

\bibitem{Bernard:1979qt}
Claude~W. Bernard.
\newblock {Gauge Zero Modes, Instanton Determinants, and QCD Calculations}.
\newblock {\em Phys. Rev. D}, 19:3013, 1979.

\bibitem{Dorey:2002ik}
Nick Dorey, Timothy~J. Hollowood, Valentin~V. Khoze, and Michael~P. Mattis.
\newblock {The Calculus of many instantons}.
\newblock {\em Phys. Rept.}, 371:231--459, 2002.

\bibitem{Tong:2005un}
David Tong.
\newblock {TASI lectures on solitons: Instantons, monopoles, vortices and
  kinks}.
\newblock In {\em {Theoretical Advanced Study Institute in Elementary Particle
  Physics}: {Many Dimensions of String Theory}}, 6 2005.

\bibitem{AlvarezGaume:1981hn}
Luis Alvarez-Gaume, Daniel~Z. Freedman, and Sunil Mukhi.
\newblock {The Background Field Method and the Ultraviolet Structure of the
  Supersymmetric Nonlinear Sigma Model}.
\newblock {\em Annals Phys.}, 134:85, 1981.

\bibitem{FrancoisMS}
Francois David.
\newblock {\em Geometry and Field Theory of Random Surfaces and Membranes},
  pages 149--209.

\bibitem{JamesSimons68}
James Simons.
\newblock Minimal varieties in riemannian manifolds.
\newblock {\em Annals of Mathematics}, 88(1):62--105, 1968.

\bibitem{AnciauxMS}
Henri Anciaux.
\newblock {\em Minimal submanifolds in pseudo-Riemannian geometry}.
\newblock 01 2010.

\bibitem{DeWittMorette:1976up}
C.~DeWitt-Morette.
\newblock {The Semiclassical Expansion}.
\newblock {\em Annals Phys.}, 97:367--399, 1976.
\newblock [Erratum: Annals Phys. 101, 682 (1976)].

\bibitem{Banerjee:2005bb}
Rabin Banerjee, Pradip Mukherjee, and Anirban Saha.
\newblock {Bosonic p-brane and A-D-M decomposition}.
\newblock {\em Phys. Rev. D}, 72:066015, 2005.

\bibitem{brink2013principles}
L.~Brink and M.~Henneaux.
\newblock {\em Principles of String Theory}.
\newblock Series of the Centro De Estudios Cient{\'\i}ficos. Springer US, 2013.

\bibitem{Hoppe_2012}
Jens Hoppe.
\newblock Relativistic membranes.
\newblock {\em Journal of Physics A: Mathematical and Theoretical},
  46(2):023001, dec 2012.

\bibitem{Schild-1977}
Alfred Schild.
\newblock Classical null strings.
\newblock {\em Physical Review D}, 16, 9 1977.

\bibitem{Karlhede-1986}
U~Karlhede, A;~Lindstrom.
\newblock The classical bosonic string in the zero tension limit.
\newblock {\em Classical and Quantum Gravity}, 3, 07 1986.

\bibitem{Amorim:1987bk}
R.~Amorim and J.~Barcelos-Neto.
\newblock {Strings With Zero Tension}.
\newblock {\em Z. Phys. C}, 38:643, 1988.

\bibitem{BarcelosNeto:1989gs}
J.~Barcelos-Neto, C.~Ramirez, and M.~Ruiz-Altaba.
\newblock {PHASE SPACE LAGRANGIANS FOR NULL SPINNING STRINGS}.
\newblock {\em Z. Phys. C}, 47:241--246, 1990.

\bibitem{J-1990}
J.~Gamboa; Cupatitzio Ramírez;~M. Ruiz-Altaba.
\newblock Null spinning strings.
\newblock {\em Nuclear Physics B}, 338, 1990.

\bibitem{Hassani:1994rf}
S.~Hassani, U.~Lindstrom, and R.~von Unge.
\newblock {Classically equivalent actions for tensionless p-branes}.
\newblock {\em Class. Quant. Grav.}, 11:L79--L85, 1994.

\bibitem{Symanzik:1981wd}
K.~Symanzik.
\newblock {Schrodinger Representation and Casimir Effect in Renormalizable
  Quantum Field Theory}.
\newblock {\em Nucl. Phys. B}, 190:1--44, 1981.

\bibitem{Luscher:1985iu}
M.~Luscher.
\newblock {SCHRODINGER REPRESENTATION IN QUANTUM FIELD THEORY}.
\newblock {\em Nucl. Phys. B}, 254:52--57, 1985.

\bibitem{Jackiw:1988sf}
R.~Jackiw.
\newblock {ANALYSIS ON INFINITE DIMENSIONAL MANIFOLDS: SCHRODINGER
  REPRESENTATION FOR QUANTIZED FIELDS}.
\newblock pages 383--445, 8 1988.

\bibitem{Jacobson:2003vx}
Ted Jacobson.
\newblock {Introduction to quantum fields in curved space-time and the Hawking
  effect}.
\newblock In {\em {School on Quantum Gravity}}, pages 39--89, 8 2003.

\bibitem{Mansfield:1993pd}
Paul Mansfield.
\newblock {Continuum strong coupling expansion of Yang-Mills theory: Quark
  confinement and infrared slavery}.
\newblock {\em Nucl. Phys. B}, 418:113--130, 1994.

\bibitem{1996NCimB.111...85H}
T.~{Horiguchi}.
\newblock {WKB approximation and renormalizability of the Wheeler-DeWitt
  equation}.
\newblock {\em Nuovo Cimento B Serie}, 111(1):85--92, January 1996.

\bibitem{grigoryan2009heat}
A.~Grigoryan.
\newblock {\em Heat Kernel and Analysis on Manifolds}.
\newblock AMS/IP studies in advanced mathematics. American Mathematical
  Society, 2009.

\bibitem{Papadimitriou:2010as}
Ioannis Papadimitriou.
\newblock {Holographic renormalization as a canonical transformation}.
\newblock {\em JHEP}, 11:014, 2010.

\bibitem{LevSa91}
Levitan and Sargsjan I.S.
\newblock {\em {Sturm—Liouville and Dirac Operators}}.
\newblock Mathematics and its Applications. Springer Netherlands, Berlin,
  Germany, 1991.

\bibitem{Levai:1989eaa}
G.~Levai.
\newblock {A Search for Shape Invariant Solvable Potentials}.
\newblock {\em J. Phys.}, A22:689--702, 1989.

\bibitem{doi:10.1142/4687}
Fred Cooper, Avinash Khare, and Uday Sukhatme.
\newblock {\em Supersymmetry in Quantum Mechanics}.
\newblock WORLD SCIENTIFIC, 2001.

\bibitem{NIST:DLMF}
{\it NIST Digital Library of Mathematical Functions}.
\newblock http://dlmf.nist.gov/, Release 1.0.26 of 2020-03-15.
\newblock F.~W.~J. Olver, A.~B. {Olde Daalhuis}, D.~W. Lozier, B.~I. Schneider,
  R.~F. Boisvert, C.~W. Clark, B.~R. Miller, B.~V. Saunders, H.~S. Cohl, and
  M.~A. McClain, eds.

\bibitem{HardyR}
G.~H. Hardy and Marcel Riesz.
\newblock {\em The general theory of Dirichlet's series}.
\newblock University Press, Cambridge {$[$}Eng.{$]$}, 1915.

\bibitem{helson1963}
Henry Helson.
\newblock {Convergent Dirichlet series}.
\newblock {\em Ark. Mat.}, 4(6):501--510, 01 1963.

\bibitem{mandelbrojt2012dirichlet}
S.~Mandelbrojt.
\newblock {\em Dirichlet Series: Principles and Methods}.
\newblock Springer Netherlands, 2012.

\bibitem{apostol1998introduction}
T.M. Apostol.
\newblock {\em Introduction to Analytic Number Theory}.
\newblock Undergraduate Texts in Mathematics. Springer New York, 1998.

\bibitem{Witten:1998qj}
Edward Witten.
\newblock {Anti-de Sitter space and holography}.
\newblock {\em Adv. Theor. Math. Phys.}, 2:253--291, 1998.

\bibitem{Freedman:1998tz}
Daniel~Z. Freedman, Samir~D. Mathur, Alec Matusis, and Leonardo Rastelli.
\newblock {Correlation functions in the CFT(d) / AdS(d+1) correspondence}.
\newblock {\em Nucl. Phys.}, B546:96--118, 1999.

\bibitem{deAlfaro:1976vlx}
Vittorio de~Alfaro, S.~Fubini, and G.~Furlan.
\newblock {Conformal Invariance in Quantum Mechanics}.
\newblock {\em Nuovo Cim.}, A34:569, 1976.

\bibitem{griffiths2017introduction}
D.J. Griffiths.
\newblock {\em Introduction to Quantum Mechanics}.
\newblock Cambridge University Press, 2017.

\bibitem{10.2307/41291944}
BRUCE~C. BERNDT and BYUNGCHAN KIM.
\newblock Asymptotic expansions of certain partial theta functions.
\newblock {\em Proceedings of the American Mathematical Society},
  139(11):3779--3788, 2011.

\bibitem{MaoRen2013}
Renrong Mao.
\newblock Some new asymptotic expansions of certain partial theta functions.
\newblock {\em The Ramanujan Journal}, 34:443--448, 08 2013.

\bibitem{Dzagier006}
Don Zagier.
\newblock {\em The Mellin transform and other useful analytic techniques}.
\newblock Springer-Verlag Berlin Heidelberg, 2006.

\bibitem{Dai:2009zza}
Wu-Sheng Dai and Mi~Xie.
\newblock {The number of eigenstates: counting function and heat kernel}.
\newblock {\em JHEP}, 02:033, 2009.

\bibitem{BransonGilkey90}
Thomas~P. Branson and Peter~B. Gilkey.
\newblock The asymptotics of the laplacian on a manifold with boundary.
\newblock {\em Communications in Partial Differential Equations},
  15(2):245--272, 1990.

\bibitem{Pang2012}
Hai Pang, Wu-Sheng Dai, and Mi~Xie.
\newblock Relation between heat kernel method and scattering spectral method.
\newblock {\em The European Physical Journal C}, 72, 05 2012.

\bibitem{Schwinger1954}
Julian Schwinger.
\newblock The theory of quantized fields. vi.
\newblock {\em Phys. Rev.}, 94:1362--1384, 06 1954.

\bibitem{Graham:2001iv}
N.~Graham, R.L. Jaffe, M.~Quandt, and H.~Weigel.
\newblock {Finite energy sum rules in potential scattering}.
\newblock {\em Annals Phys.}, 293:240, 2001.

\bibitem{doi:10.1119/1.2165248}
Andrew~M. Essin and David~J. Griffiths.
\newblock Quantum mechanics of the $\frac{1}{x^2}$ potential.
\newblock {\em American Journal of Physics}, 74(2):109--117, 2006.

\bibitem{Mukhanov:2007zz}
Viatcheslav Mukhanov and Sergei Winitzki.
\newblock {\em {Introduction to quantum effects in gravity}}.
\newblock Cambridge University Press, 6 2007.

\bibitem{Bastianelli:2005rc}
Fiorenzo Bastianelli.
\newblock {Path integrals in curved space and the worldline formalism}.
\newblock In {\em {8th International Conference on Path Integrals from Quantum
  Information to Cosmology}}, 8 2005.

\bibitem{Ishizeki:2011bf}
Riei Ishizeki, Martin Kruczenski, and Sannah Ziama.
\newblock {Notes on Euclidean Wilson loops and Riemann Theta functions}.
\newblock {\em Phys. Rev. D}, 85:106004, 2012.

\bibitem{Kruczenski:2013bsa}
Martin Kruczenski and Sannah Ziama.
\newblock {Wilson loops and Riemann theta functions II}.
\newblock {\em JHEP}, 05:037, 2014.

\bibitem{Kruczenski:2014bla}
Martin Kruczenski.
\newblock {Wilson loops and minimal area surfaces in hyperbolic space}.
\newblock {\em JHEP}, 11:065, 2014.

\bibitem{Irrgang:2015txa}
Andrew Irrgang and Martin Kruczenski.
\newblock {Euclidean Wilson loops and minimal area surfaces in lorentzian
  AdS$_{3}$}.
\newblock {\em JHEP}, 12:083, 2015.

\bibitem{Huang:2016atz}
Changyu Huang, Yifei He, and Martin Kruczenski.
\newblock {Minimal area surfaces dual to Wilson loops and the Mathieu
  equation}.
\newblock {\em JHEP}, 08:088, 2016.

\bibitem{He:2017zsk}
Yifei He and Martin Kruczenski.
\newblock {Minimal area surfaces in $AdS_3$ through integrability}.
\newblock {\em J. Phys. A}, 50(49):495401, 2017.

\bibitem{He:2017cwd}
Yifei He, Changyu Huang, and Martin Kruczenski.
\newblock {Minimal area surfaces in AdS\_{n+1} and Wilson loops}.
\newblock {\em JHEP}, 02:027, 2018.

\bibitem{Cooke:2018obg}
Michael Cooke, Amit Dekel, Nadav Drukker, Diego Trancanelli, and Edoardo
  Vescovi.
\newblock {Deformations of the circular Wilson loop and spectral
  (in)dependence}.
\newblock {\em JHEP}, 01:076, 2019.

\bibitem{10.2307/1999231}
M.~Do Carmo and M.~Dajczer.
\newblock Rotation hypersurfaces in spaces of constant curvature.
\newblock {\em Transactions of the American Mathematical Society},
  277(2):685--709, 1983.

\bibitem{Tuz1992}
Alexey Tuzhilin.
\newblock Morse-type indices of two-dimensional minimal surfaces in r3 and h3.
\newblock {\em Mathematics of The Ussr-izvestiya}, 38:575--598, 06 1992.

\bibitem{Wang2016}
Biao Wang.
\newblock Stability of catenoids and helicoids in hyperbolic space.
\newblock {\em Asian Journal of Mathematics}, 23, 08 2016.

\bibitem{Dubrovin1981}
B~A Dubrovin.
\newblock Theta functions and non-linear equations.
\newblock {\em Russian Mathematical Surveys}, 36(2):11--92, apr 1981.

\bibitem{kalla:tel-00622289}
Caroline Kalla.
\newblock {\em {Fay's identity in the theory of integrable systems}}.
\newblock Theses, {Universit{\'e} de Bourgogne}, June 2011.

\bibitem{Pastras:2016vqu}
Georgios Pastras.
\newblock {Static elliptic minimal surfaces in AdS$_4$}.
\newblock {\em Eur. Phys. J. C}, 77(11):797, 2017.

\bibitem{Anselme_Dossa_2020}
Finagnon~Anselme Dossa.
\newblock Pauli hamiltonian for a spin one-half particle carrying a non-abelian
  charge in the presence of non-abelian external fields.
\newblock {\em {EPL} (Europhysics Letters)}, 131(2):21002, aug 2020.

\bibitem{HKESUBEP}
Ovidiu Calin, Der-Chen Chang, Kenro Furutani, and C.~Iwasaki.
\newblock {\em Heat Kernels for Elliptic and Sub-elliptic Operators: Methods
  and Techniques}, pages 1--431.
\newblock 01 2011.

\bibitem{benedetti1992lectures}
R.~Benedetti, R.B.C. Petronio, and C.~Petronio.
\newblock {\em Lectures on Hyperbolic Geometry}.
\newblock Universitext (Berlin. Print). Springer Berlin Heidelberg, 1992.

\bibitem{Drukker:2000rr}
Nadav Drukker and David~J. Gross.
\newblock {An Exact prediction of N=4 SUSYM theory for string theory}.
\newblock {\em J. Math. Phys.}, 42:2896--2914, 2001.

\bibitem{Nesterenko:2002ng}
V.V. Nesterenko, I.G. Pirozhenko, and J.~Dittrich.
\newblock {Nonsmoothness of the boundary and the relevant heat kernel
  coefficients}.
\newblock {\em Class. Quant. Grav.}, 20:431--456, 2003.

\end{thebibliography}

\end{document}